\documentclass[preprint,review,12pt]{elsarticle}
\usepackage{subfigure}
\usepackage{textcomp}
\usepackage{gensymb}
\usepackage{graphicx}
\usepackage{amssymb}
\usepackage{amsmath}
\usepackage{caption}
\usepackage{url} 
\usepackage{xr}
\makeatletter
\newcommand*{\addFileDependency}[1]{
  \typeout{(#1)}
  \@addtofilelist{#1}
  \IfFileExists{#1}{}{\typeout{No file #1.}}
}
\makeatother

\newcommand*{\myexternaldocument}[1]{
    \externaldocument{#1}
    \addFileDependency{#1.tex}
    \addFileDependency{#1.aux}
}
\myexternaldocument{SI}

\usepackage{float}
\restylefloat{table}

\journal{Acta Materialia}

\begin{document}
\begin{frontmatter}


\title{Grain Boundary Properties of Elemental Metals}



\author[UCSD]{Hui Zheng \fnref{cor2}}
\author[UCSD]{Xiang-Guo Li \fnref{cor2}}
\author[UCSD]{Richard Tran}
\author[UCSD]{Chi Chen}
\fntext[cor2]{These authors contributed equally}

\author[LBNL]{Matthew Horton}
\author[LBNL]{Donny Winston}
\author[LBNL,UCB]{Kristin Aslaug Persson}

\author[UCSD]{Shyue Ping Ong \corref{cor1}}
\ead{ongsp@eng.ucsd.edu}
\cortext[cor1]{Corresponding author}

\address[UCSD]{Department of NanoEngineering, University of California San Diego, 9500 Gilman Dr, Mail Code 0448, La Jolla, CA 92093-0448, United States}
\address[LBNL]{Environmental Energy Technologies Division, Lawrence Berkeley National Laboratory, Berkeley,  CA 94720, United States}
\address[UCB]{Department of Materials Science \& Engineering, University of California Berkeley, Berkeley,  CA 94720-1760, United States}

\begin{abstract}
The structure and energy of grain boundaries (GBs) are essential for predicting the properties of polycrystalline materials. In this work, we use high-throughput density functional theory calculations workflow to construct the Grain Boundary Database (GBDB), the largest database of DFT-computed grain boundary properties  to date. The database currently encompasses 327 GBs of 58 elemental metals, including 10 common twist or symmetric tilt GBs for body-centered cubic (bcc) and face-centered cubic (fcc) systems and the $\Sigma$7 [0001] twist GB for hexagonal close-packed (hcp) systems. In particular, we demonstrate a novel scaled-structural template approach for HT GB calculations, which reduces the computational cost of converging GB structures by a factor of $\sim 3-6$. The grain boundary energies and work of separation are rigorously validated against previous experimental and computational data. Using this large GB dataset, we develop an improved predictive model for the GB energy of different elements based on the cohesive energy and shear modulus. The open GBDB represent a significant step forward in the availability of first principles GB properties, which we believe would help guide the future design of polycrystalline materials.
\end{abstract}


\end{frontmatter}


\section{Introduction}
\label{S:1}

The majority of engineering materials are polycrystals, comprising a large number of grains whose interfaces form grain boundaries (GBs). The GB character distribution (GBCD)\cite{Watanabe1992}, i.e., the type and frequency of GBs present, strongly affects a material's mechanical properties\cite{Rohrer2011d, Watanabe2011a} such as hardness\cite{Hu2018b}, brittleness\cite{Watanabe1999a, Zheng2018}, creep-strength\cite{LEHOCKEY1997}, corrosion resistance\cite{Shi2017}, fatigue strength\cite{Kobayashi2008}, and weldability\cite{Lehockey1998}. For instance, intergranular fracture is the primary origin of severe brittleness and fatigue failure, and GBs are the preferential sites for the nucleation and propagation of fatigue cracks \cite{Watanabe1999a,Was1990}. Manipulating the GBCD through various processing techniques is a common pathway to improving the mechanical properties of structural metals and alloys. \cite{Watanabe1999a,Kobayashi2008,Watanabe2011,KRUPP2003213, Pineau2015}.  

The GBCD of a material is related to the relative GB formation energies\cite{Rohrer2010a}. Typically, the lower the formation energy for a particular type of GB (otherwise simply known as the GB energy or $\gamma_{GB}$), the greater its prevalence in the polycrystal\cite{Rohrer2010, Zheng2018}. A variety of experimental techniques (e.g., thermal groove, orientation imaging microscopy) have been applied to investigate $\gamma_{GB}$ , but the data sets were limited due to the difficulty of measuring accurate $\gamma_{GB}$ \cite{HASSON1972115, BARMAK20061059, GJOSTEIN1959319, McLean1973, Chan1986, Miura1994, Skidmore2004a}. Recently Rohrer et al. have developed a high-throughput (HT) experimental method to measure $\gamma_{GB}$ for large ensembles of GBs by inversely correlating it with the statistical abundance of GB types present in the polycrystal \cite{Rohrer2010a, Rohrer2011e, Amouyal2005a}. This method has been applied to fcc Ni \cite{Li2009}, Ni-based alloys \cite{Rohrer2010a}, W thin film\cite{Liu2013a}, ferrite (mainly bcc Fe) \cite{BELADI20131404}, austenitic steel (mainly fcc Fe)\cite{BELADI2014281} and hcp Ti \cite{Kelly2016}. Such HT studies have significantly increased the available experimental data for $\gamma_{GB}$\cite{Rohrer2010, Li2009}. However, this statistical approach suffers from a strong dependence of the uncertainty in the measured $\gamma_{GB}$ on the frequency of observed GBs, leading to unreliable measurements for GBs of lower frequency. Furthermore, the method yields relative, rather than absolute, $\gamma_{GB}$. 

Computationally, there have been many investigations of $\gamma_{GB}$ using both empirical and first principles methods. Studies using empirical interatomic potentials (IAPs) such as the embedded atom method (EAM)\cite{Wolf1989e, Wolf1989c, Wolf1989b} and Lennard-Jones\cite{Wolf1989e, Wolf1989c} potentials are typically limited to a few elemental systems belonging to a specific crystal prototype (e.g., fcc or bcc), but cover a broad range of GB types\cite{Wolf1990d, Wolf1990e,Olmsted2009,Holm2010a,Tschopp2015,Ratanaphan2015b}. The reason is because the fitting of sufficiently accurate IAPs is a relatively complex and resource-intensive process, but once fitted, it is inexpensive to use the IAP to compute many GB structures comprising thousands or even millions of atoms. For instance, \citet{Olmsted2009,Holm2010a, Holm2011a} have calculated $\gamma_{GB}$ for 388 distinct GBs of fcc Ni, Al, Au, and Cu using EAM and found that GB energies in different elements are strongly correlated. For bcc metals, \citet{Ratanaphan2015b} have computed the energies of 408 distinct GBs in bcc Fe and Mo ranging from $\Sigma$3 to $\Sigma$323. Their results show that GB energies are influenced more by GB plane orientation than by lattice misorientation or lattice coincidence. 

With computing advances, calculations of $\gamma_{GB}$ using accurate, but expensive first-principles methods such as density functional theory (DFT) have become increasingly common. In contrast to IAP-based studies, DFT studies tend to be broader in chemical scope but narrow in the range of GB structures studied (typically limited to low $\Sigma$ GB models of hundreds of atoms). This is due to the good transferability, but high computational expense, of first principles methods. For example, \citet{Scheiber2016f} have computed 14 types of GBs for W, Mo and Fe using DFT, while \citet{Wang2018i} have calculated 11 types of low sigma ($\Sigma$ $<$ 13) symmetrical tilt GBs and 2 twist GBs for bcc Fe. \citet{Bean2016a} have also used DFT calculations to verify a small subset of symmetric tilt GB structures acquired from EAM calculations in Cu and Ni systems. 

In this work, we report the development of the Grain Boundary DataBase (GBDB), a comprehensive database for GB properties ($\gamma_{GB}$, work of separation $W_{sep}$) for a broad range of low-index GB structures (tilt and twist) for fcc, bcc, and hcp elemental metals using high-throughput DFT calculations. At the time of writing, this GBDB contains data on 327 GB structures for 58 elements, with more GB types and elements continually being added. This GBDB has been made available via the Materials Project and its Application Programming Interface\cite{Ong2013, Ong2015g}, together with a user-friendly web application called Crystal Toolkit for the generation of GB structures. A critical enabler to the construction of the GBDB is an innovative lattice scaling approach, which substantially lowers the computational effort in performing GB calculations for similar crystal types across different elements. Finally, we rigorously validate the GBDB against prior experimental and computed data, and using this large dataset, develop an efficient model for predicting $\gamma_{GB}$ for different elements.

\section{Methods}
\subsection{Grain boundary model generation}

\begin{figure}[htp]
\centering\includegraphics[width=0.85\linewidth]{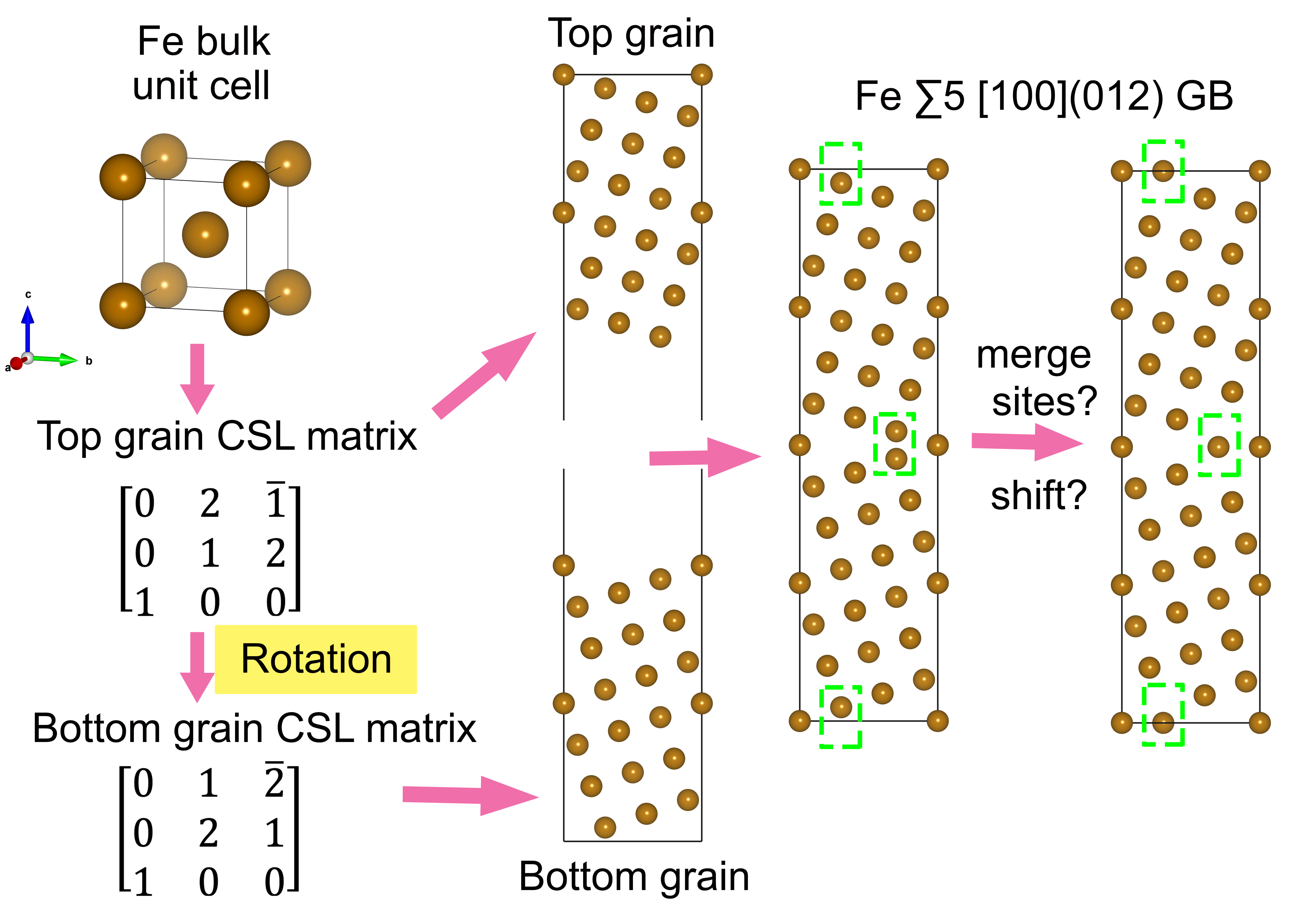}
\caption{\label{fig:grain_boundary_generation} Grain boundary generation process}
\end{figure}

Figure \ref{fig:grain_boundary_generation} shows the schematic of the GB model generation algorithm, which is based on the coincident-site lattice (CSL) method \cite{Grimmer:a28679}. For two grains misoriented by a rotation angle about a rotation axis, the superposition of the two crystals result in coincident sites forming a sublattice of the two crystal lattices, i.e., a CSL. An important parameter characterizing the CSL is the $\Sigma$ value, defined as the ratio of the unit cell volume of the CSL to the volume of the generating bulk cell. A grain boundary can be completely and unambiguously described by five macroscopic degrees of freedom (DOFs)\cite{Lejcek2010a}, e.g. $\Sigma5$ $36.87\degree/[100](031)$. Three DOFs describe the mutual misorientations between two adjoining grains, two of which define the rotation axis (two DOFs, e.g. [100]) and one of which defines the rotation angle, e.g. $36.87\degree$. The remaining two DOFs describe the GB plane, e.g. (031). The steps in the algorithm are as follows:
\begin{itemize}
    \item Starting from the unit cell (primitive or conventional cell) with lattice type of cubic, tetragonal, orthorhombic, hexagonal or rhombohedral, a series of lattice vector transformations is performed to create an unit cell of CSL with the a and b lattice vectors parallel to the input GB plane.
    \item Two grains are created and rotated relative to each other based on the inputs (rotation axis and angle, expansion times of the CSL unit cell along $c$ direction).
    \item The two grains are then stacked to form the periodic GB structure. The relative shifts between the two grains along the $a$, $b$ and $c$ directions can be adjusted. 
    \item Finally, sites that are too close to each other based on a distance tolerance set by the user are merged. 
\end{itemize}

The above algorithm is implemented in the open-source Python Materials Genomics (pymatgen) materials analysis library\cite{Ong2013}, together with methods for finding all sigma values and their corresponding rotation angles for any given input structure and rotation axis. A user-friendly graphical user interface to the algorithm is also available on Materials Project website Crystal Toolkit application (https://materialsproject.org/\#apps/xtaltoolkit).

\subsection{GB property computation}

The grain boundary energy ($\gamma_{GB}$) is defined by the following expression:

\begin{equation}
\label{eq:gb_energy}
\gamma_{GB} = \frac{E_{GB}-n_{GB}E_{bulk}}{2A_{GB}}
\end{equation}

where $E_{GB}$ and $n_{GB}$ are the total energy and number of atoms of the grain boundary structure, respectively, $A_{GB}$ is the cross-sectional area of the GB, $E_{bulk}$ is the energy per atom of the bulk, and the factor of 2 in the denominator accounts for the two grain boundaries in the GB model.

Another GB property of interest is the work of separation $W_{sep}$, which is a measure of the energy required to cleave the GB into the free surfaces and is correlated to the fracture toughness\cite{MOLLER20141, Coffman2008b, GRUJICIC1997341, GUMBSCH199972}. $W_{sep}$ is given by the following expression: 
\begin{equation}
\label{eq:work_of_separation}
W_{sep} = 2\gamma_{surf} - \gamma_{GB}
\end{equation}

where $\gamma_{surf}$ is the corresponding surface energy for the facet $(hkl)$ formed by cleaving the GB. Previously, some of the current authors have already constructed a comprehensive database of the surface energies of the elements\cite{Tran2016a}, which are used in this work in the computation of $W_{sep}$.

\subsection{DFT computations}

All DFT energy calculations were performed using the Vienna Ab initio Simulation Package (VASP) \cite{Kohn1965} with the projector augmented wave (PAW) \cite{Kresse1996a, Blochl1994} method. The exchange-correlation effects were modeled using the Perdew-Berke-Ernzerhof (PBE) \cite{Perdew1996} generalized gradient approximation (GGA) functional. The plane wave energy cutoff is 400 eV, and $k$-point grids of 30 \AA$^{-1}$ and 45 \AA$^{-1}$ in each lattice direction were used for relaxation and single-point energy calculations, respectively. The energies and atomic forces of all calculations were converged within $10^{-4}$ eV and 0.02 eV \AA$^{-1}$. Through a series of convergence tests, it was determined that a thickness of at least 25 \AA ~along the direction normal to the GB plane is sufficient to minimize periodic interactions between the two grain boundaries, such that $\gamma_{GB}$ is converged to within 0.02 J m$^{-2}$. 

\subsection{Scope of Data}

\begin{figure}[htp]
\centering\includegraphics[width=\textwidth]{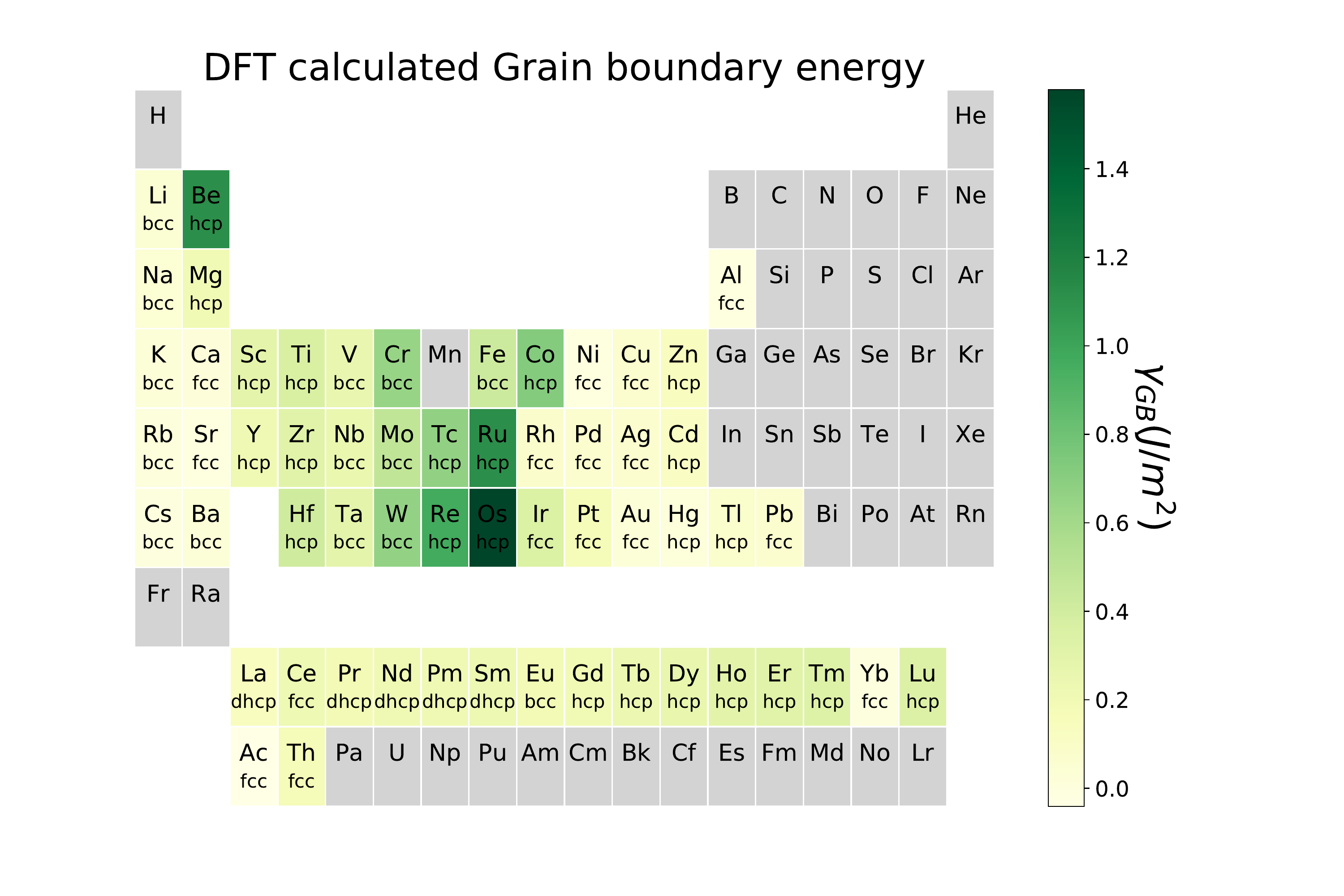}
\caption{\label{fig:GB_periodic_table} DFT calculated grain boundary energy. For bcc and fcc, the lowest $\gamma_{GB}$ types, i.e., $\Sigma$3[110](112) for bcc and $\Sigma$3[111](111) for fcc are plotted. For hcp, and double-hcp (dhcp) elements, $\Sigma$7(0001) GBs are chosen to be represented in this periodic table heatmap.}
\end{figure}

Our database covers a total of 58 elements (see Figure \ref{fig:GB_periodic_table}), with 10 GB types for fcc and bcc and one GB type for hcp and double-hcp (dhcp) elements (see Table \ref{table:gb_types}), with a total of 327 GB structures. We limit the GB types in this study with the following criteria:
\begin{enumerate}
\item $\Sigma$ $<$ 10
\item Maximum Miller index (MMI) of rotation axis $\leq$ 1
\item MMI of grain boundary plane $\leq$ 3.
\item All tilt GBs are symmetric.
\end{enumerate}

\begin{table}[H]
\centering
\caption{\label{table:gb_types} Grain boundary types calculated in this work}
\begin{tabular}{llllllll}
\hline \hline
Sigma & type  & R-axis     & R-angle & GB-plane            & Join-plane          & bcc atoms & fcc atoms \\ \hline
 \multicolumn{8}{c}{bcc or fcc} \\ \hline

3     & tilt  & {[}110{]}  & 109.47  & (1 $\bar1$ $\bar2$) & ($\bar1$ 1 $\bar2$) & 24        & 46        \\ \hline
3     & tilt  & {[}111{]}  & 180      & (1 $\bar1$ 0)       & (0 1 $\bar1$)        & 48        & 56        \\ \hline
3     & twist & {[}111{]}  & 60      & (1 1 1)             & (1 1 1)             & 48        & 24        \\ \hline
5     & tilt  & {[}100{]}  & 36.87   & (0 $\bar1$ $\bar2$) & (0 2 1)               & 38        & 38        \\ \hline
5     & tilt  & {[}100{]}  & 53.13   & (0 $\bar1$ $\bar3$) & (0 3 1)               & 40        & 58        \\ \hline
5     & twist & {[}100{]}  & 36.87   & (1 0 0)             & (1 0 0)               & 80        & 80        \\ \hline
7     & twist & {[}111{]}  & 38.21   & (1 1 1)               & (1 1 1)               & 168       & 84        \\ \hline
7     & tilt  & {[}111{]}  & 38.21   & (1 $\bar3$ 2) & ($\bar2$ 3 $\bar1$) & 54        & 54        \\ \hline
9     & twist & {[}110{]}  & 38.94   & (1 1 0)               & (1 1 0)               & 126       & 180       \\ \hline
9     & tilt  & {[}110{]}  & 38.94   & (2 $\bar2$ $\bar1$) & (2 $\bar2$ 1)        & 70        & 70        \\ \hline
 \multicolumn{6}{c}{hcp/dhcp}  & hcp/dhcp atoms   \\ \hline
7     & twist & {[}0001{]} & 21.79   & (0 0 0 1)              & (0 0 0 1)              & 112  \\ \hline
\end{tabular}
\end{table}

\section{Results}

\subsection{Benchmarking}

A major bottleneck to calculations of GBs is that the large system sizes combined with difficult convergence of atomic positions, especially close to the GB region, render such computations relatively expensive compared to bulk crystal calculations. To accelerate such computations, a fundamental hypothesis explored in this work is that similar crystal structures (e.g., bcc, fcc, or hcp) lead to similar low-energy GB configurations.

To find the most favorable initial configurations of GBs, we applied rigid body translation \cite{Tschopp2015} of two grains to each type of twist GBs by performing a series of static calculations for each translation vector. The translation unit step length along the \textbf{\textit{c}} direction is in increments of 10\% of the lattice parameter of the conventional unit cell. While along the basal directions (\textbf{\textit{a}} and \textbf{\textit{b}}), the translation unit step is in increments of 5\% - 10\% of the basal lattice vectors (\textbf{\textit{a}} and \textbf{\textit{b}}) of the GB structure. For symmetric tilt GBs, atoms at the interface that are less than 70\% of the bulk interatomic distance apart are merged. We find that the most favorable initial configurations are identical for crystals of the same prototype (see Figure S1 and S2). 

\begin{figure}[htp]
\centering\includegraphics[width=0.85\linewidth]{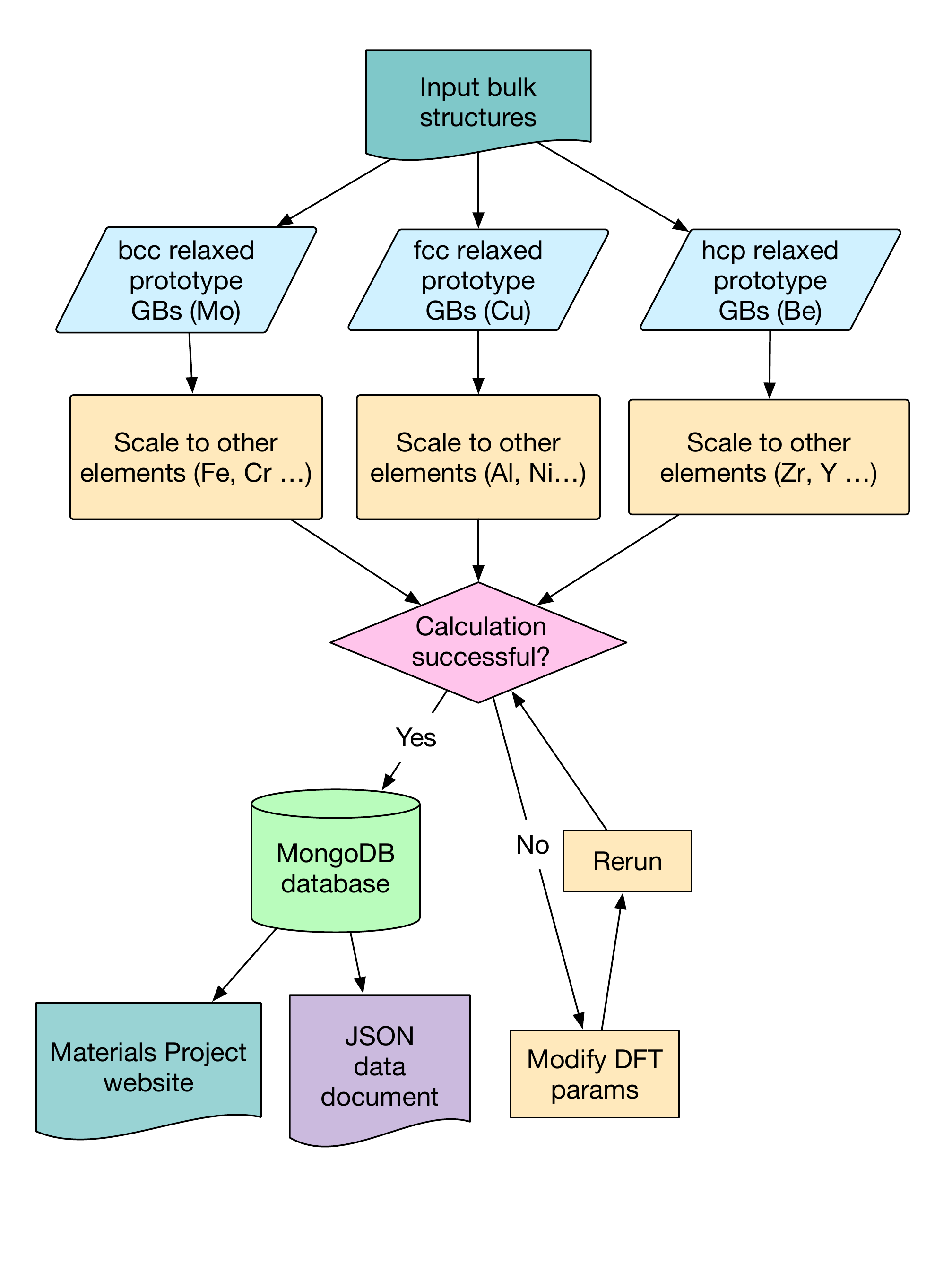}
\caption{\label{fig:workflow} High-throughput computational workflow for elemental grain boundaries.}
\end{figure}

Based on these results, we have developed a high-throughput workflow for GB calculations using the Atomate software package \cite{Ong2013,Mathew2017,Jain2015b}, as shown in Figure \ref{fig:workflow}. For each structural prototype (bcc, fcc, hcp and dhcp), we first compute a series of fully-relaxed GB templates for all the GB types investigated in this work (see Table \ref{table:gb_types}), using Mo, Cu and Be/La as the templates for bcc, fcc and hcp/dhcp structures, respectively. Initial structures for GB computations of each element M are then created from these GB templates by applying a scaling factor of $\frac{a_M}{a_{prototype}}$ to the template GB lattice constants for all materials, where $a_M$ and $a_{prototype}$ are the bulk lattice parameters of the metal M and prototype element respectively. No scaling is applied for Zn and Cd, which are hcp elements with c/a ratios (1.986 and 1.915, respectively) that deviate substantially from the ideal ratio of 1.633, and their GB structures were generated directly from the bulk structure. A full relaxation is then performed on the scaled GBs. The use of the scaled GB templates significantly reduces the computational resources for the most time-consuming structural relaxation step by a factor of $\sim 3-6$, with higher speed-ups for GBs with larger number of atoms and GBs that are very different from bulk (Table \ref{table:CPU_time}). More accurate static calculations with denser $k$-point meshes were then performed to obtain the final total energy of the GB structures. The results were then automatically inserted into a MongoDB document-based database.

\begin{table}[H]
\centering
\caption{CPU hours for GB relaxation with and without the use of scaled prototype templates.}
\begin{tabular}{llcccc}
\hline \hline
 &  &   & \multicolumn{2}{c}{CPU hours}  \\
Element & GB type & \# of atoms &  No template & With template & Speed up \\ \hline
bcc-Ba &  $\Sigma$3(111) & 48 &  2560.00 &   716.80 & 3.57\\ \hline 
bcc-Fe &  $\Sigma$9(110) & 126 &  2340.00 & 508.33 & 4.60 \\ \hline
fcc-Sr &  $\Sigma$5(100) & 80 &  2128.05 & 344.29 & 6.18 \\ \hline
fcc-Ag &  $\Sigma$5(013) & 80 &  97.67 & 97.55 & 1.00 \\ \hline
hcp-Ti &  $\Sigma$7(0001) & 112 &  24.28 & 13.94 &  1.74 \\ \hline
dhcp-Nd &  $\Sigma$7(0001) & 112 &  218.39 & 59.08 &  3.70 \\ \hline
\end{tabular}
\label{table:CPU_time}
\end{table}

\subsection{Grain boundary energies}

\begin{figure}[htp]
\centering\includegraphics[width=\linewidth]{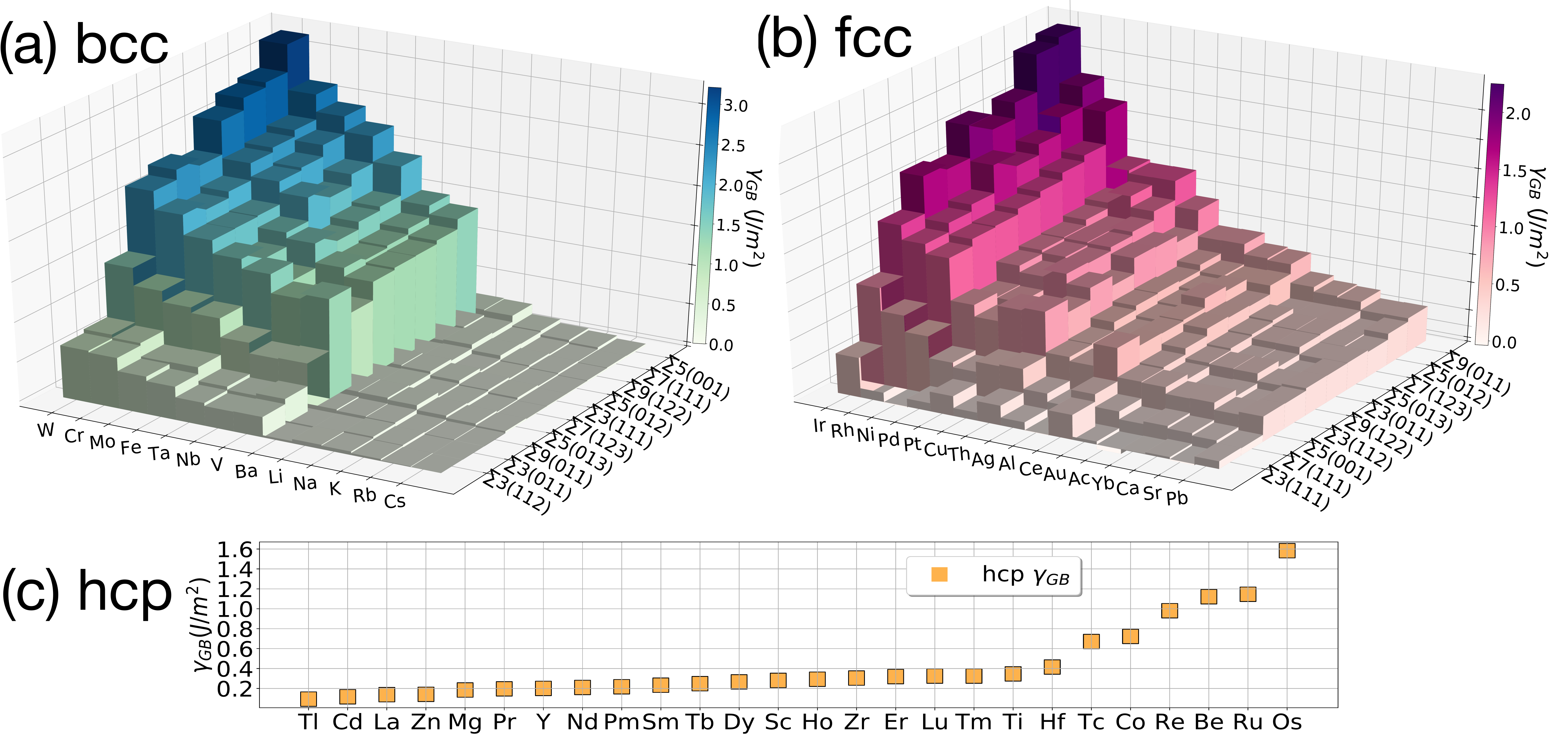}
\caption{\label{fig:3D_colormap_bar} Grain boundary energy $\gamma_{GB}$ distribution for (a) bcc, (b) fcc, and (c) hcp/dhcp elemental metals, sorted by increasing $\gamma_{GB}$.}
\end{figure}

Figure \ref{fig:3D_colormap_bar} shows the distribution of $\gamma_{GB}$ for bcc, fcc, and hcp elements. All values are tabulated in Table S1 and S2 for reference. For bcc elements (Figure \ref{fig:3D_colormap_bar}a), we can observe a substantial jump in $\gamma_{GB}$ from alkali/alkaline earth metals to transition metals; the $\gamma_{GB}$ for alkali and alkaline earth metals are less than 0.3 $Jm^{-2}$, while those for the transition metals are at least four times higher. $\gamma_{GB}$ for fcc elements follows a similar trend but with a more gradual increase (see Figure \ref{fig:3D_colormap_bar}b). Group VIII elements have high $\gamma_{GB}$ while group IB, IIA, and IIB elements have relatively low $\gamma_{GB}$. 

Figure \ref{fig:3D_colormap_bar}c shows the $\gamma_{GB}$ distribution for hcp/dhcp $\Sigma$7(0001) grain boundaries. For transition metals, we observe that $\gamma_{GB}$ peaks at groups VIIB and VIII (Tc, Co, Re, Ru, and Os). All the rare earth and group IIA elements have lower GB energies than the transition metals with the exception of Be, which has a much higher GB energy. The rare earth elements show a gradual increase in $\gamma_{GB}$ as group number increases. 

The $\gamma_{GB}$ distribution across different GB types varies with the crystal type. The two coherent twin boundaries, $\Sigma$3(111) for fcc and $\Sigma3(112)$ for bcc, have the lowest $\gamma_{GB}$ within the respective crystal prototypes. GBs terminated by the most atomically-dense planes ((111) for fcc and (110) for bcc) have lower $\gamma_{GB}$ than other planes in general. Consequently, the fcc $\Sigma$ 7 (111) and bcc $\Sigma$3(011) GBs correspond to the second lowest $\gamma_{GB}$ for fcc and bcc, respectively. This is in agreement with both previous atomistic calculations\cite{Holm2010a, Ratanaphan2015b, Wolf1991, Wolf1989c, Wolf1989b} and experimental results\cite{Rohrer2010,Holm2011a, Saylor2004a, Liu2013a}. For example, it has been observed experimentally that the most frequently observed grain boundary for fcc Ni and Al is the $\Sigma$3(111) twin boundary, and other GBs terminated with the (111) plane also have a high population\cite{Li2009, Saylor2004a}. For bcc metals, our data shows that the $\Sigma$3(112) symmetric tilt GB (twin) has the lowest energy, which agrees with experiments performed in bcc W thin films\cite{Liu2013a} with nanoscale grain sizes and bcc ferritic/interstitial free steel\cite{BELADI20131404, Beladi2013a}. 

\begin{figure}[H]
\subfigure[]
{\label{fig:compare_gb_e_with_other_DFT}\includegraphics[width=0.3\textwidth]{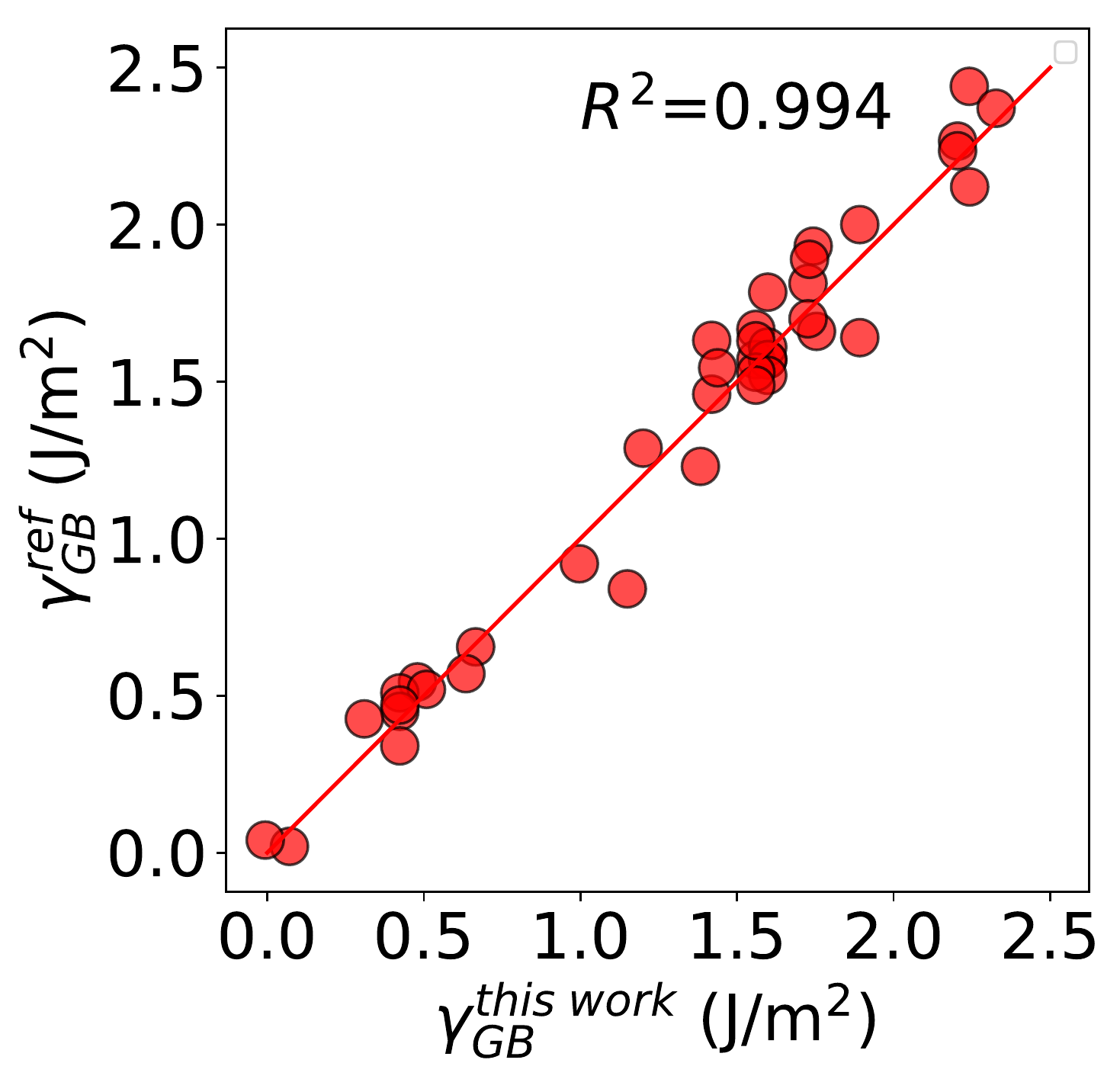}}
\subfigure[]
{\label{fig:Egb_Mo_DFT_EAM_SNAP}\includegraphics[width=0.3\textwidth]{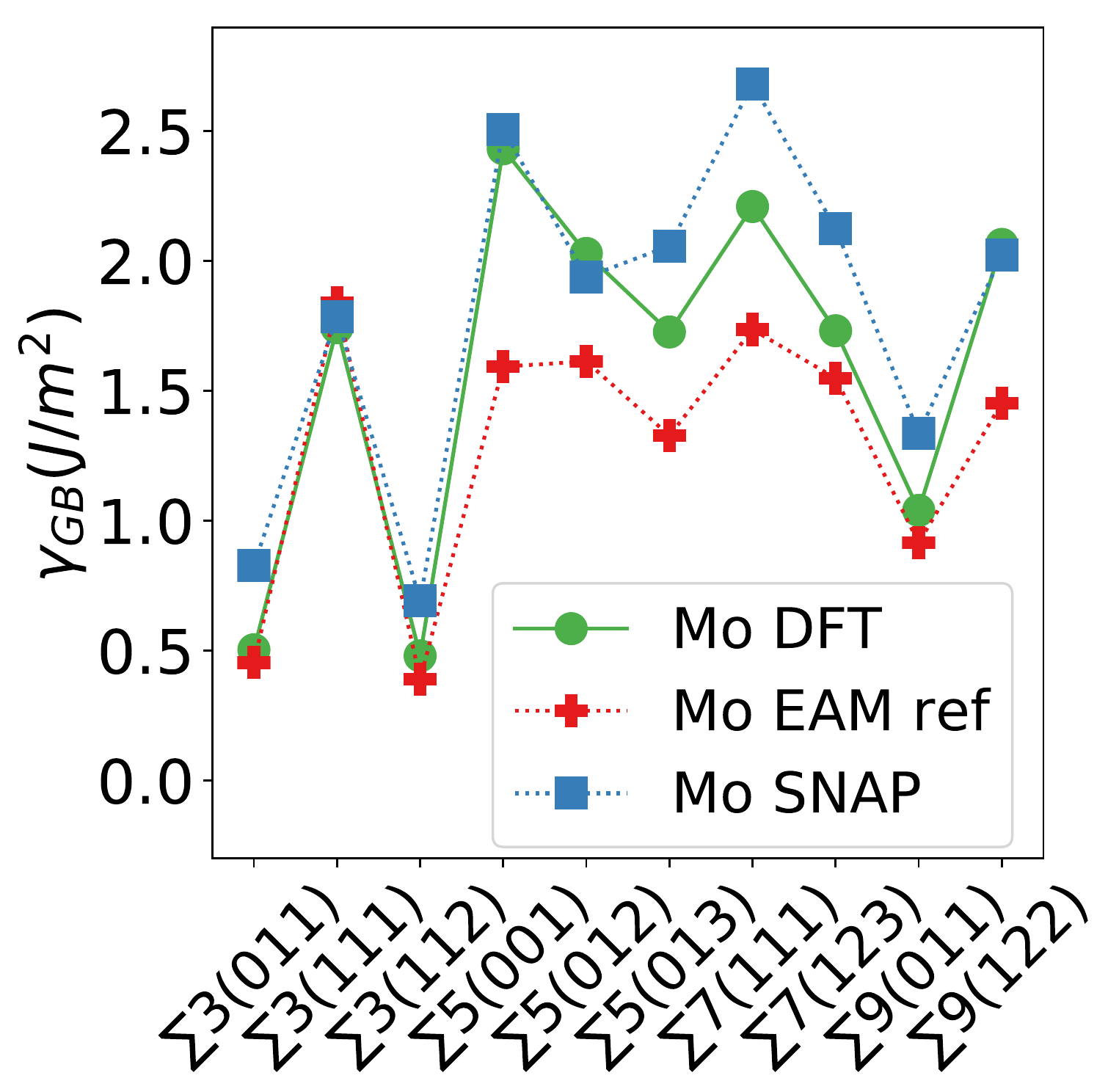}}
\subfigure[]
{\label{fig:Egb_Ni_DFT_EAM_SNAP}\includegraphics[width=0.31\textwidth]{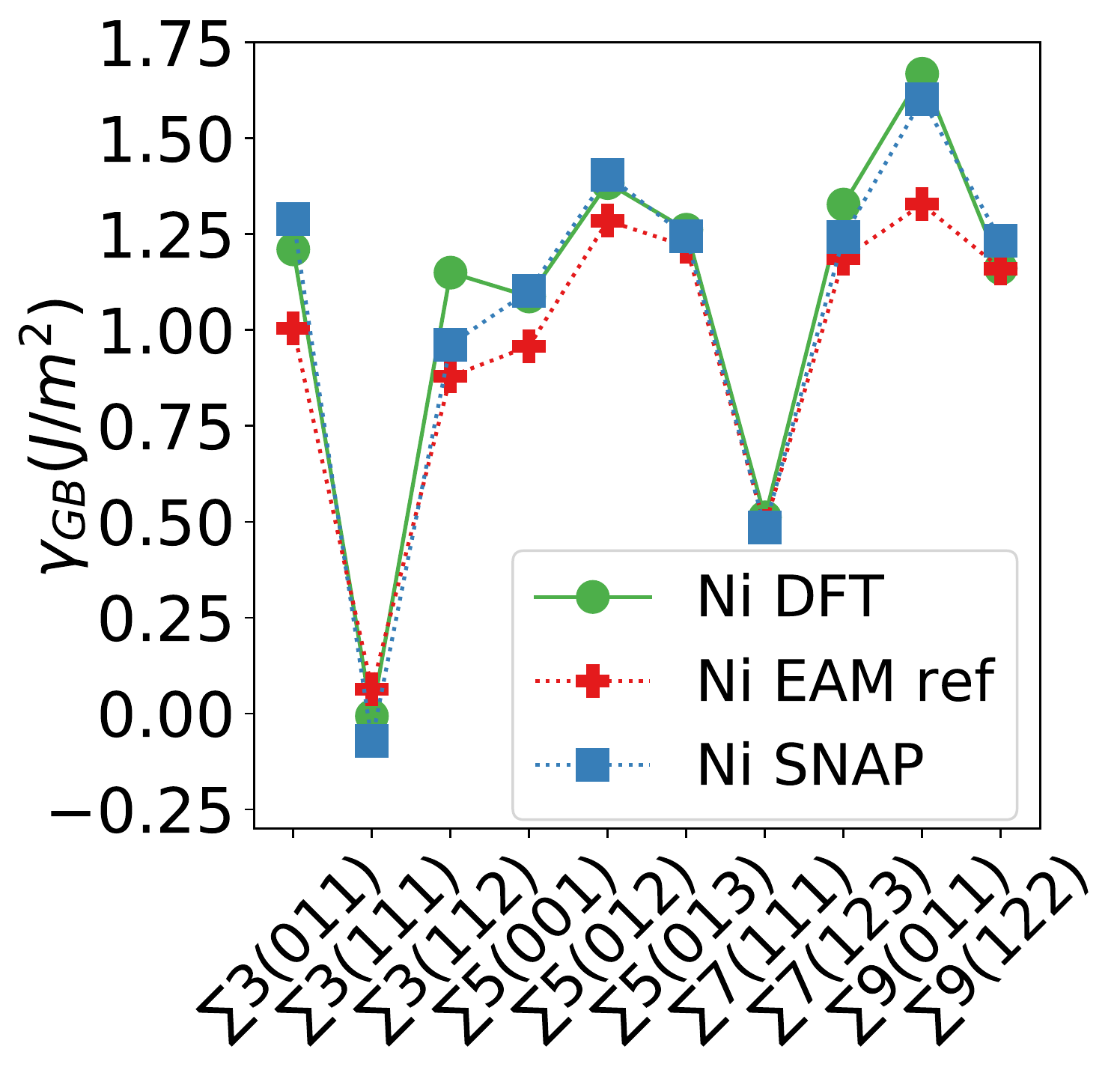}}
\subfigure[]
{\label{fig:bcc_Fe_dft_vs_exp_lnMRD}\includegraphics[width=0.31\textwidth]{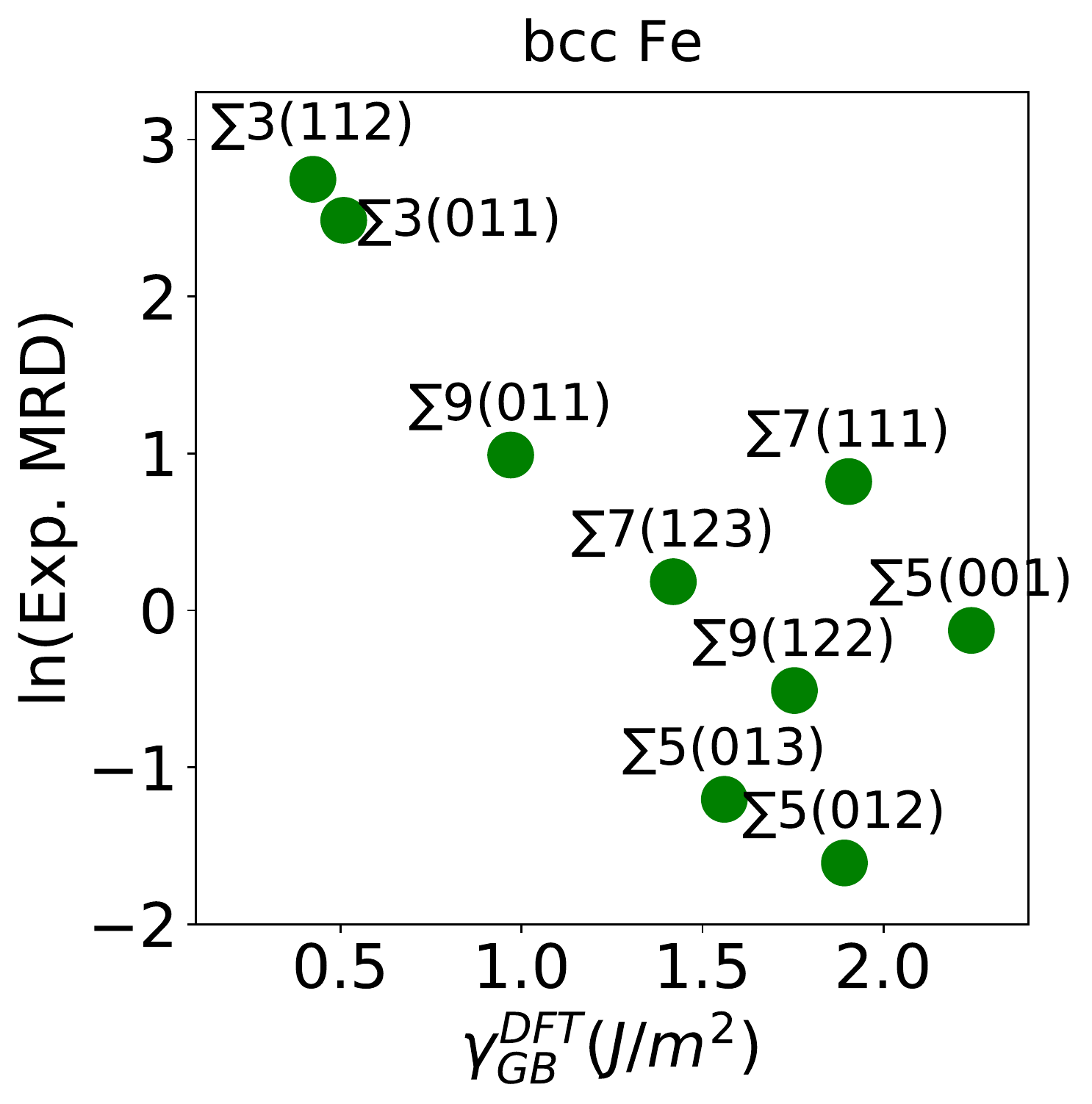}}
\subfigure[]
{\label{fig:fcc_Al_dft_vs_exp_lnMRD}\includegraphics[width=0.3\textwidth]{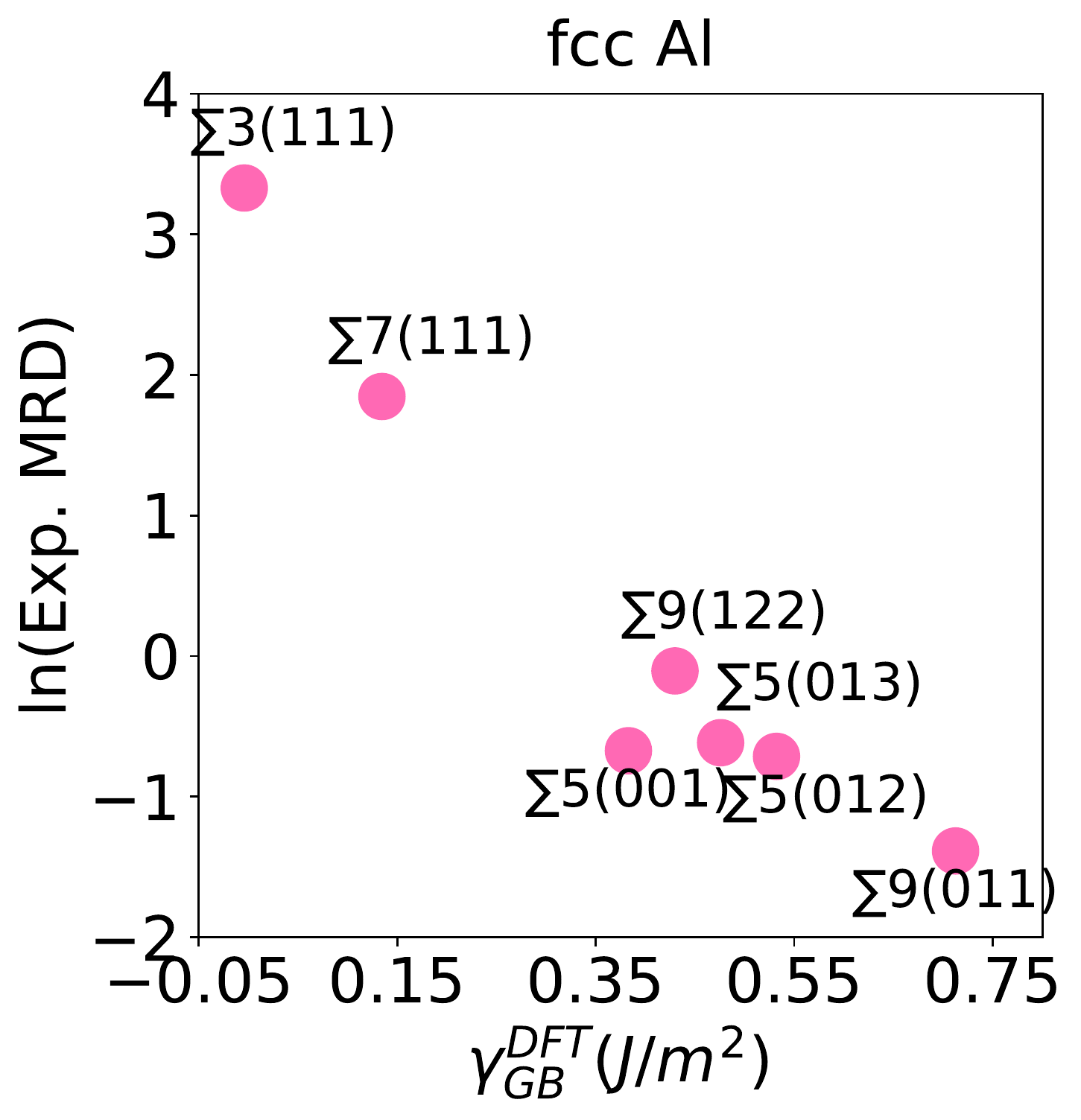}}
\subfigure[]
{\label{fig:fcc_Ni_exp_smoothed_gbe_vs_DFT}\includegraphics[width=0.32\textwidth]{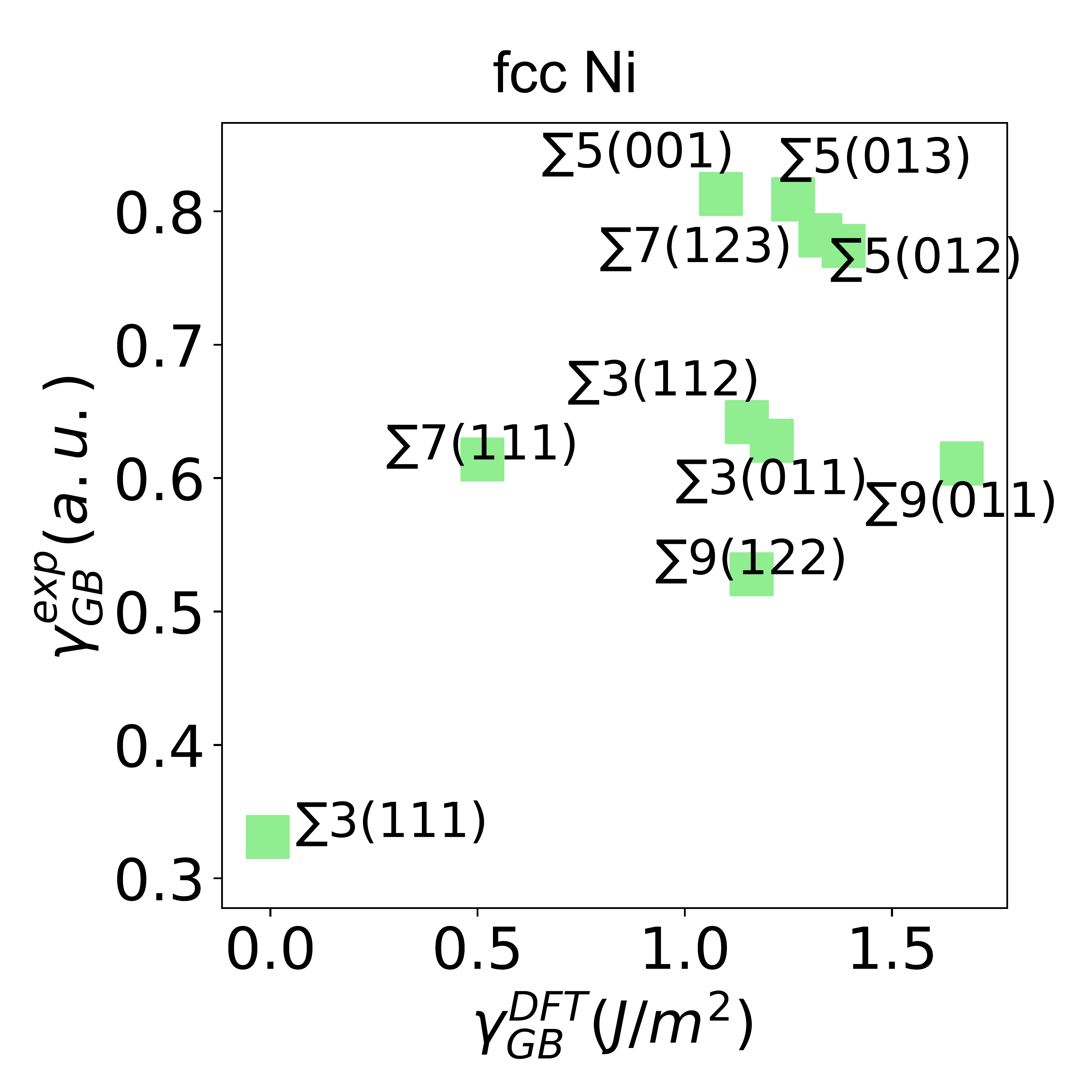}}

\caption{\label{fig:validation} Comparison of $\gamma_{GB}$ between this work and (a) previous DFT values; (b, c) EAM \cite{Olmsted2009, Ratanaphan2015b} and SNAP\cite{Chen2017, Li2018j} values. (d), (e) and (f) compare our the calculated $\gamma_{GB}$ of bcc Fe, fcc Al, and fcc Ni with experimentally measured MRD\cite{Beladi2013a, Saylor2004a} and $\gamma_{GB}$\cite{Li2009}. }
\end{figure}

Figure \ref{fig:validation} shows the validation of our computed $\gamma_{GB}$ with previous DFT calculations \cite{Scheiber2016f, Wang2018i,Bhattacharya2013,Wachowicz2010, Gao2009, Du2011, Cak2008, Ochs2000a,Bean2016a, Wright1994} , atomistic calculations\cite{PLIMPTON19951} using machine-learned spectral neighbor analysis potentials (SNAP)\cite{Chen2017, Li2018j} and the embedded atom method (EAM)\cite{Ratanaphan2015b, Olmsted2009, Holm2011a}, and experimental data\cite{Li2009, Rohrer2010, Beladi2013a, Liu2013a}. From Figure \ref{fig:validation}a and Table S5 we may observe that our computed $\gamma_{GB}$ are in excellent agreement with previous DFT values, with a $R^{2}$ close to unity and a very small standard error of 0.013 J/m$^2$. Similarly, we find good agreement between the calculated $\gamma_{GB}$ for different GBs of Mo and Ni with those computed using the state-of-the-art SNAP models\cite{Chen2017, Li2018j}, while the EAM predicted GB energies\cite{Olmsted2009, Ratanaphan2015b} are substantially underestimated as shown in Figure \ref{fig:validation}(b) and (c). For bcc Mo, values of $\gamma_{GB}$ using SNAP are slightly larger than most DFT values with the exception of the $\Sigma5 (012)$ GB where SNAP slightly underestimates DFT values. For fcc Ni, the $\gamma_{GB}$ values of both EAM and SNAP are consistent with our DFT values, further supporting the conclusion that EAM performs better in fcc systems than bcc systems\cite{Li2018j}.

Figures \ref{fig:validation}(d) and (e) plot the natural log of the experimentally measured multiples of random distribution (MRD), i.e. the experimental average population of GBs, against the DFT calculated grain boundary energy ($\gamma^{DFT}_{GB}$) for Fe and Al, respectively. We observe a negative correlation between the $\ln(MRD)$ and $\gamma_{GB}$ similar to that reported previously for Ni\cite{Li2009}. Figure \ref{fig:validation} (f) plots the experimental grain boundary energy ($\gamma_{GB}^{exp}$)\cite{Li2009} against our $\gamma_{GB}^{DFT}$ values for Ni. All values of $\gamma_{GB}^{exp}$ are derived from a statistical average of the MRD and given in arbitrary units. We also observe a general positive correlation between $\gamma_{GB}^{exp}$ and $\gamma_{GB}^{DFT}$.

\subsection{Work of separation}

\begin{figure}[htp]
\subfigure[bcc $\Sigma$3(110)]
{\label{fig:bcc_s3_011_Wsep}\includegraphics[width=0.3\textwidth]{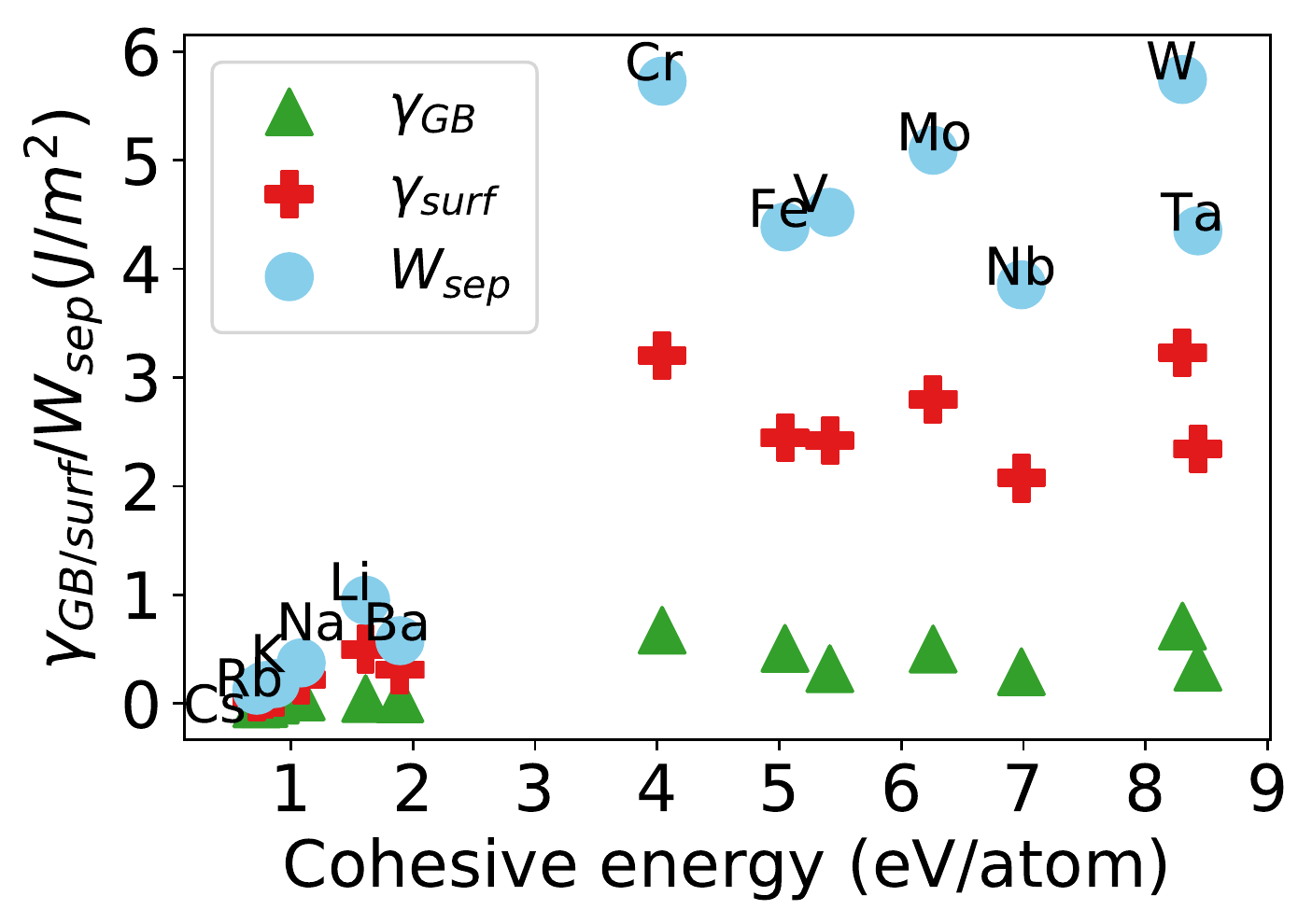}}
\subfigure[fcc $\Sigma$3(111)]
{\label{fig:fcc_s3_111_Wsep}\includegraphics[width=0.3\textwidth]{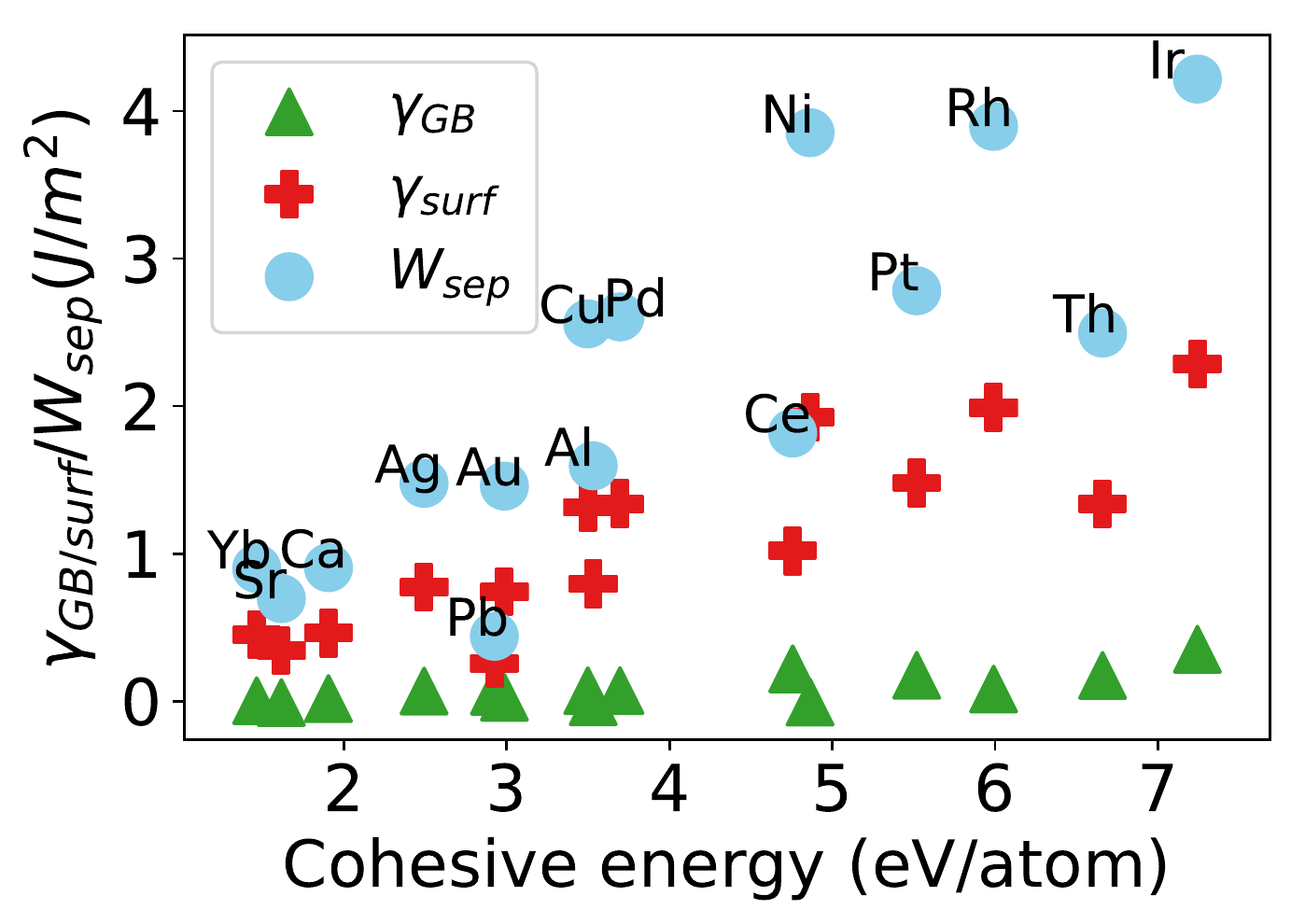}}
\subfigure[hcp $\Sigma$7(0001)]
{\label{fig:hcp_s7_0001_Wsep}\includegraphics[width=0.3\textwidth]{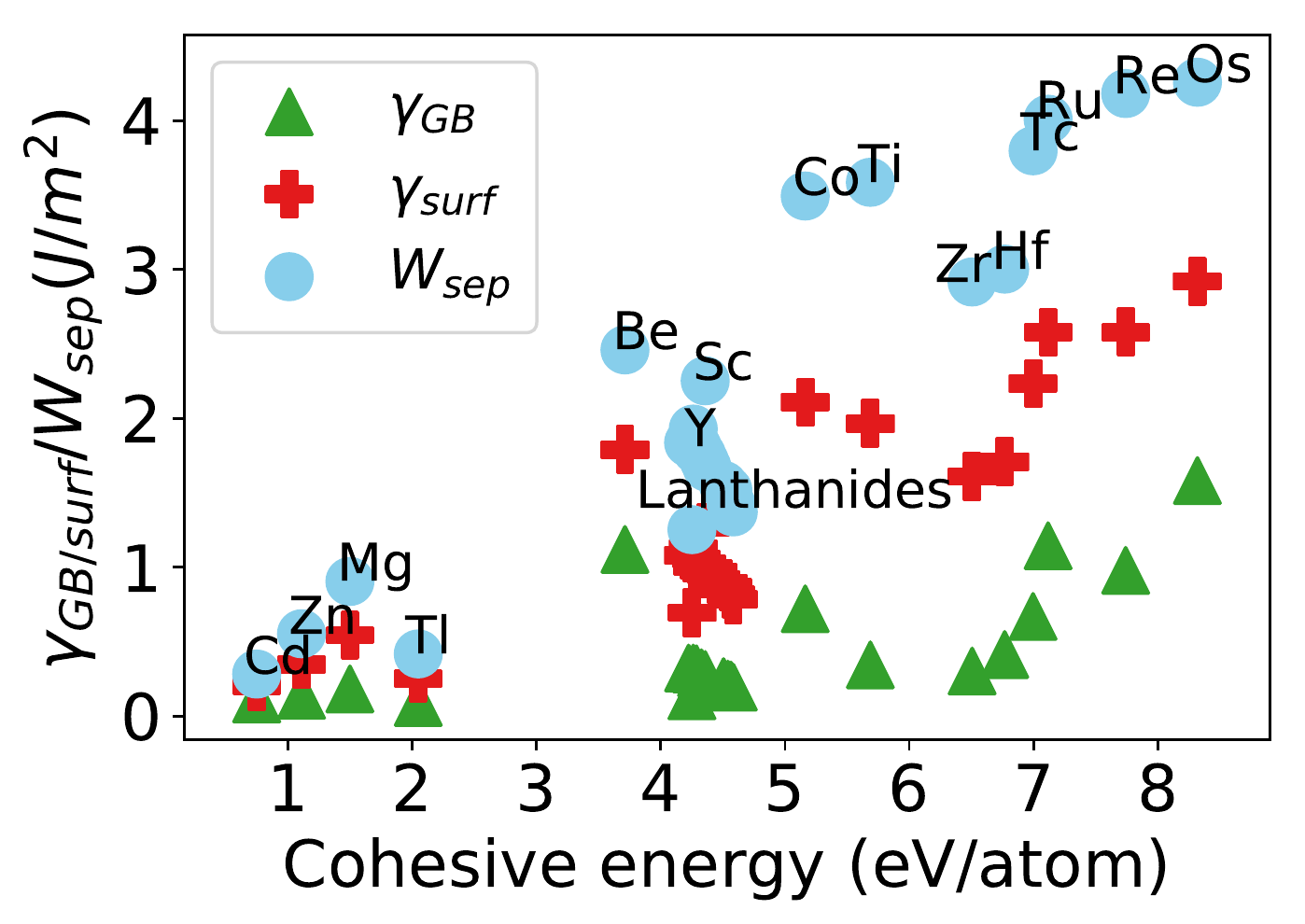}}
\caption{\label{fig:Wsep_Esurf_Egb} Comparison between surface energy ($\gamma_{surf}$), grain boundary energy ($\gamma_{GB}$) and work of separation ($W_{sep}$) for (a) bcc $\Sigma$3(110), (b) fcc $\Sigma$3(111) and (c) hcp $\Sigma$7(0001) GBs, plotted in order of ascending cohesive energy $E_{coh}$ of the element.}
\end{figure}

The thermodynamic threshold energy for GB fracture, or work of separation ($W_{sep}$), can be defined as the difference between the surface energy and GB energy as shown in equation (\ref{eq:work_of_separation}) . Since the formation of surfaces and GBs both relate to bond breaking and distortion, we expect grain boundary energy $\gamma_{GB}$, surface energy $\gamma_{surf}$ and work of separation $W_{sep}$ to be positively correlated with cohesive energy. This is demonstrated in Figure \ref{fig:Wsep_Esurf_Egb} for bcc $\Sigma$3(110), fcc $\Sigma$3(111) and hcp $\Sigma$7(0001) GBs. The values of $W_{sep}$ for all other GB types are provided in Table S3 and S4. This positive correlation is in agreement with previous bond breaking arguments \cite{Wolf1990d, Wolf1990e}. In general, the variation in anisotropic surface energies across different surfaces is smaller compared to the GB energy variation across different types of GBs as shown in Figure S3. As such, we can expect a negative correlation between GB energy and work of separation as shown in Figure S4.

\begin{figure}[htp]
\centering\includegraphics[width=\linewidth]{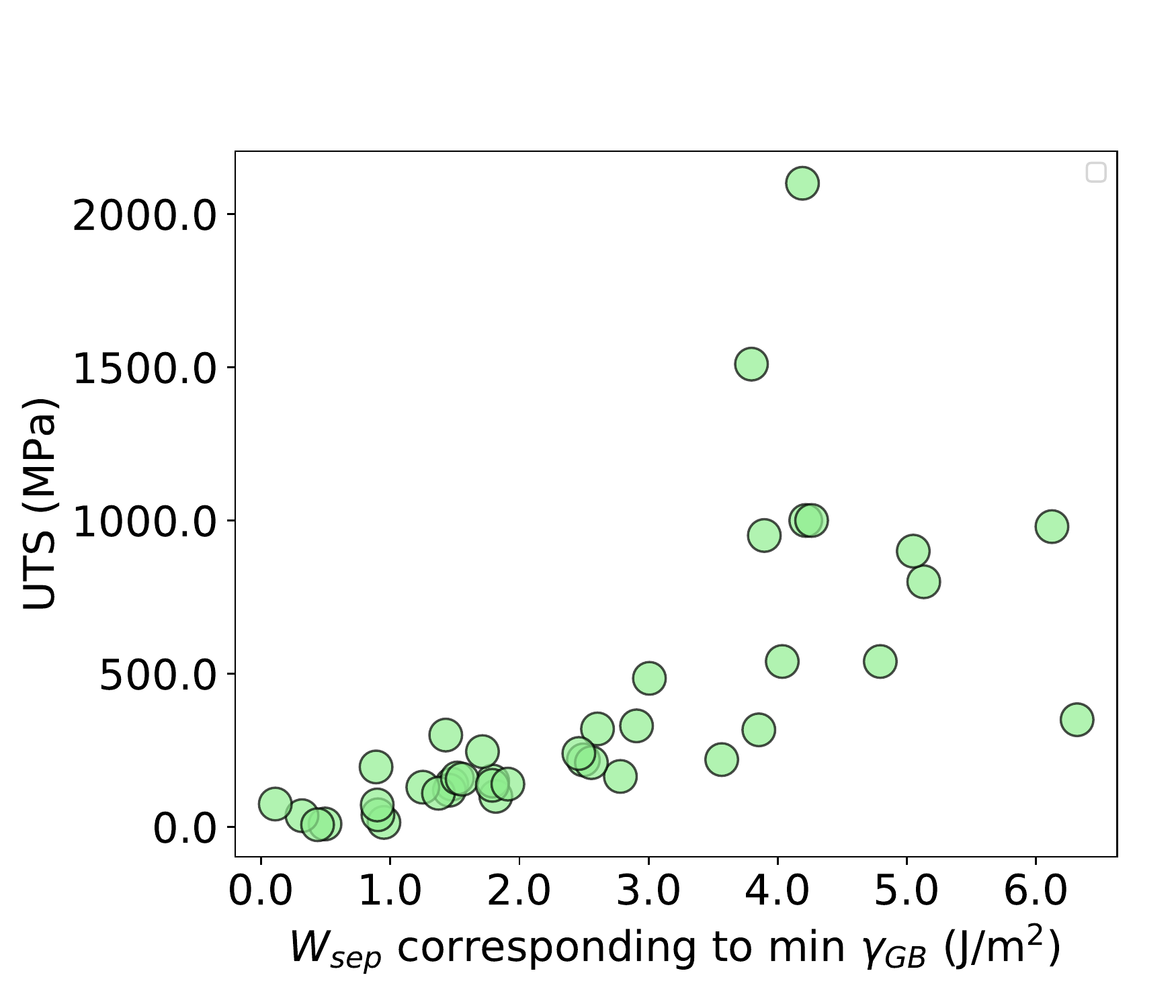}
\caption{\label{fig:UTS_Wsep} Relationship between calculated work of separation $W_{sep}$ for GB with lowest $\gamma_{GB}$ and experimentally measured ultimate tensile strength (UTS) \cite{web:exp_UTS}.}
\end{figure}

Figure \ref{fig:UTS_Wsep} plots the experimentally measured ultimate tensile strength (UTS)\cite{web:exp_UTS} against the calculated $W_{sep}$ for the GB with lowest $\gamma_{GB}$, i.e., the likely dominant GB type. A general positive relationship is observed between $W_{sep}$ and UTS, as expected. The non-monotonic relationship may be due to the different processing methods (e.g., annealing, heat treatment, cold-worked) that can significantly affect micro-structure, and hence measured UTS.

\subsection{Multiple linear regression model for $\gamma_{GB}$}
Using the extensive set of computed $\gamma_{GB}$, we have developed a multiple linear regression (MLR) model for $\gamma_{GB}$ for each GB type by fitting to the following equation:

\begin{equation}
\label{eq:Egb_fitting}
\widehat{\gamma_{GB}} = \beta_{1}E_{coh}a_0^{-2} + \beta_{2}G \cdot a_0
\end{equation}

where $\widehat{\gamma_{GB}}$ is the fitted grain boundary energy, $E_{coh}$ is the cohesive energy, $a_0$ is the lattice parameter of corresponding conventional bulk cell (\AA), and $G$ is the shear modulus (Jm$^{-3}$)\cite{DeJong2015}. This model choice is an amalgamation of models proposed in previous works. Ratanaphan et al. have found that the $\gamma_{GB}$ of bcc Fe and Mo are strongly correlated with the cohesive energy  ($E_{coh}$)\cite{Ratanaphan2015b}. Previous EAM-based GB databases have also found that $\gamma_{GB}$ for fcc metals such as Al, Au, Cu and Ni are strongly correlated to the Voigt average shear modulus ($C_{44}$)\cite{Holm2010a, Olmsted2009}. Furthermore, the Read-Shockley dislocation model\cite{PhysRev.78.275} treats GBs with small misorientation angles as an array of dislocations whose energy is proportional to a shear modulus. In essence, the $E_{coh}a_0^{-2}$ term in equation (\ref{eq:Egb_fitting})  accounts for the contribution of broken bonds to $\gamma_{GB}$, while the $G \cdot a_0$ term accounts for the contributions from distorted (stretched, compressed) bonds. Both terms have been scaled by powers of the lattice constant such that the coefficients $\beta_{1}$ and $\beta_{2}$ are dimensionless.

\begin{figure}[htp]
\centering\includegraphics[width=0.85\linewidth]{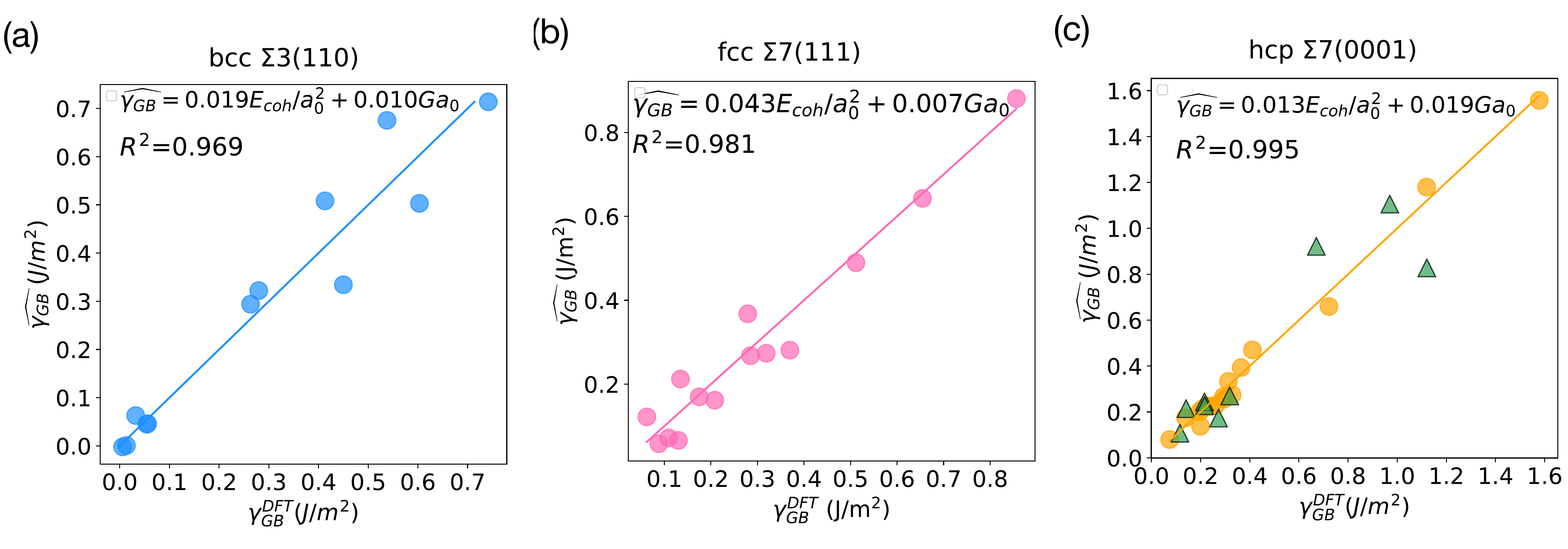}
\caption{\label{fig:fitting} Multiple linear regression models for the (a) bcc $\Sigma3(110)$, (b) fcc $\Sigma7(111)$, and (c) hcp $\Sigma7(0001)$ GBs.}
\end{figure}

Figure \ref{fig:fitting} shows the fitting results for three GB types (see Figure S5 and S6 for the remaining GB types). In general, the MLR models exhibit good predictive accuracy across all GB types, with $R^2 > 0.9$. We note that each GB type has different fitted values of the dimensionless coefficients $\beta_{1}$ and $\beta_{2}$ due to different contributions from bond breaking and bond distortion. We provide an example to show the predictive ability of our linear regression model. In Figure \ref{fig:fitting}(c), the orange circles are the data points used to build the MLR model, and the green triangles are a ``test set'' of elemental GBs. It can be seen that the performance on the ``test set'' is similar to that of the training set. We show that these results hold for all the GB structures computed in this work, and we believe it will hold for GB structures of larger $\Sigma$ values for which the model GB structure can contain many more atoms and hence are more expensive to compute. The implication of these results is that a predictive MLR model can potentially be constructed using a smaller set of elements with a range of $E_{coh}$ and $G$, and use to extrapolate to other elements.

\section{Conclusion}

The GBDB is, to the best of our knowledge, the most comprehensive database of DFT computed GB energies and work of separation to date, spanning 10 different of GB types, including both tilt and twist GBs, across 58 types of metals. This GBDB has been rigorously validated with previous computational values as well as experimental observations of the GBCD\cite{Saylor2004a,Rohrer2010a, Rohrer2010, Rohrer2011e, Liu2013a, Beladi2013a, Ratanaphan2015b}. The linear regression model provides an inexpensive estimate for the GB energy of elemental metals using cohesive energy and shear modulus. 

\section{Acknowledgement}

This work is supported by the Materials Project, funded by the U.S. Department of Energy, Office of Science, Office of Basic Energy Sciences, Materials Sciences and Engineering Division under Contract no. DE-AC02-05-CH11231: Materials Project program KC23MP. The authors also acknowledge computational resources provided by the National Energy Research Scientific Computing Centre (NERSC), the Triton Shared Computing Cluster (TSCC) at the University of California, San Diego, and the Extreme Science and Engineering Discovery Environment (XSEDE) supported by National Science Foundation under grant no. ACI-1053575. The authors thank Professor Gregory S. Rohrer from Carnegie Mellon University for providing the exact values of the experimental grain boundary energies of Ni from ref \cite{Li2009}.


%% References

\section{References}

\bibliographystyle{model1-num-names}
\bibliography{GBDB.bib}

\begin{thebibliography}{10}
\expandafter\ifx\csname natexlab\endcsname\relax\def\natexlab#1{#1}\fi
\providecommand{\bibinfo}[2]{#2}
\ifx\xfnm\relax \def\xfnm[#1]{\unskip,\space#1}\fi
%Type = Article
\bibitem[{Scheiber et~al.(2016)Scheiber, Pippan, Puschnig, and
  Romaner}]{Scheiber2016f}
\bibinfo{author}{D.~Scheiber}, \bibinfo{author}{R.~Pippan},
  \bibinfo{author}{P.~Puschnig}, \bibinfo{author}{L.~Romaner},
\newblock \bibinfo{title}{{Ab initio calculations of grain boundaries in bcc
  metals}},
\newblock \bibinfo{journal}{Modelling and Simulation in Materials Science and
  Engineering} \bibinfo{volume}{24} (\bibinfo{year}{2016})
  \bibinfo{pages}{035013}.
%Type = Article
\bibitem[{Wang et~al.(2018)Wang, Madsen, and Drautz}]{Wang2018i}
\bibinfo{author}{J.~Wang}, \bibinfo{author}{G.~K.~H. Madsen},
  \bibinfo{author}{R.~Drautz},
\newblock \bibinfo{title}{Grain boundaries in bcc-fe: a density-functional
  theory and tight-binding study},
\newblock \bibinfo{journal}{Modelling and Simulation in Materials Science and
  Engineering} \bibinfo{volume}{26} (\bibinfo{year}{2018})
  \bibinfo{pages}{025008}.
%Type = Article
\bibitem[{Bhattacharya et~al.(2013)Bhattacharya, Tanaka, Shiihara, and
  Kohyama}]{Bhattacharya2013}
\bibinfo{author}{S.~K. Bhattacharya}, \bibinfo{author}{S.~Tanaka},
  \bibinfo{author}{Y.~Shiihara}, \bibinfo{author}{M.~Kohyama},
\newblock \bibinfo{title}{Ab initiostudy of symmetrical tilt grain boundaries
  in bcc fe: structural units, magnetic moments, interfacial bonding, local
  energy and local stress},
\newblock \bibinfo{journal}{Journal of Physics: Condensed Matter}
  \bibinfo{volume}{25} (\bibinfo{year}{2013}) \bibinfo{pages}{135004}.
%Type = Article
\bibitem[{Wachowicz et~al.(2010)Wachowicz, Ossowski, and
  Kiejna}]{Wachowicz2010}
\bibinfo{author}{E.~Wachowicz}, \bibinfo{author}{T.~Ossowski},
  \bibinfo{author}{A.~Kiejna},
\newblock \bibinfo{title}{{Cohesive and magnetic properties of grain boundaries
  in bcc Fe with Cr additions}},
\newblock \bibinfo{journal}{Phys. Rev. B} \bibinfo{volume}{81}
  (\bibinfo{year}{2010}) \bibinfo{pages}{94104}.
%Type = Article
\bibitem[{Gao et~al.(2009)Gao, Fu, Samaras, Sch{\"{a}}ublin, Victoria, and
  Hoffelner}]{Gao2009}
\bibinfo{author}{N.~Gao}, \bibinfo{author}{C.~C. Fu},
  \bibinfo{author}{M.~Samaras}, \bibinfo{author}{R.~Sch{\"{a}}ublin},
  \bibinfo{author}{M.~Victoria}, \bibinfo{author}{W.~Hoffelner},
\newblock \bibinfo{title}{{Multiscale modelling of bi-crystal grain boundaries
  in bcc iron}},
\newblock \bibinfo{journal}{Journal of Nuclear Materials} \bibinfo{volume}{385}
  (\bibinfo{year}{2009}) \bibinfo{pages}{262--267}.
%Type = Article
\bibitem[{Du et~al.(2011)Du, Ismer, Jutta, Hickel, Neugebauer, and
  Drautz}]{Du2011}
\bibinfo{author}{Y.~A. Du}, \bibinfo{author}{L.~Ismer},
  \bibinfo{author}{R.~Jutta}, \bibinfo{author}{T.~Hickel},
  \bibinfo{author}{J.~Neugebauer}, \bibinfo{author}{R.~Drautz},
\newblock \bibinfo{title}{{First-principles study on the interaction of H
  interstitials with grain boundaries in $\alpha$ and $\gamma$-Fe}},
\newblock \bibinfo{journal}{Phys. Rev. B} \bibinfo{volume}{84}
  (\bibinfo{year}{2011}) \bibinfo{pages}{144121}.
%Type = Article
\bibitem[{{\v{C}}{\'{a}}k et~al.(2008){\v{C}}{\'{a}}k, {\v{S}}ob, and
  Hafner}]{Cak2008}
\bibinfo{author}{M.~{\v{C}}{\'{a}}k}, \bibinfo{author}{M.~{\v{S}}ob},
  \bibinfo{author}{J.~Hafner},
\newblock \bibinfo{title}{{First-principles study of magnetism at grain
  boundaries in iron and nickel}},
\newblock \bibinfo{journal}{Physical Review B - Condensed Matter and Materials
  Physics} \bibinfo{volume}{78} (\bibinfo{year}{2008}) \bibinfo{pages}{1--10}.
%Type = Article
\bibitem[{Ochs et~al.(2000)Ochs, Els{\"{a}}sser, Mrovec, Vitek, Belak, and
  Moriarty}]{Ochs2000a}
\bibinfo{author}{T.~Ochs}, \bibinfo{author}{C.~Els{\"{a}}sser},
  \bibinfo{author}{M.~Mrovec}, \bibinfo{author}{V.~Vitek},
  \bibinfo{author}{J.~Belak}, \bibinfo{author}{J.~A. Moriarty},
\newblock \bibinfo{title}{{Symmetrical tilt grain boundaries in bcc transition
  metals: Comparison of semiempirical with ab-initio total-energy
  calculations}},
\newblock \bibinfo{journal}{Philosophical Magazine A: Physics of Condensed
  Matter, Structure, Defects and Mechanical Properties} \bibinfo{volume}{80}
  (\bibinfo{year}{2000}) \bibinfo{pages}{2405--2423}.
%Type = Article
\bibitem[{Bean and McKenna(2016)}]{Bean2016a}
\bibinfo{author}{J.~J. Bean}, \bibinfo{author}{K.~P. McKenna},
\newblock \bibinfo{title}{{Origin of differences in the excess volume of copper
  and nickel grain boundaries}},
\newblock \bibinfo{journal}{Acta Materialia} \bibinfo{volume}{110}
  (\bibinfo{year}{2016}) \bibinfo{pages}{246--257}.
%Type = Article
\bibitem[{Wright and Atlas(1994)}]{Wright1994}
\bibinfo{author}{A.~F. Wright}, \bibinfo{author}{S.~R. Atlas},
\newblock \bibinfo{title}{{Density-functional calculations for grain boundaries
  in aluminum}},
\newblock \bibinfo{journal}{Phys. Rev. B} \bibinfo{volume}{50}
  (\bibinfo{year}{1994}) \bibinfo{pages}{15248--15260}.

\end{thebibliography}


\begin{thebibliography}{74}
\expandafter\ifx\csname natexlab\endcsname\relax\def\natexlab#1{#1}\fi
\providecommand{\bibinfo}[2]{#2}
\ifx\xfnm\relax \def\xfnm[#1]{\unskip,\space#1}\fi
%Type = Article
\bibitem[{Watanabe(1992)}]{Watanabe1992}
\bibinfo{author}{T.~Watanabe},
\newblock \bibinfo{title}{{The importance of grain boundary character
  distribution (GBCD) to recrystallization, grain growth and texture}},
\newblock \bibinfo{journal}{Scripta Metallurgica et Materialia}
  \bibinfo{volume}{27} (\bibinfo{year}{1992}) \bibinfo{pages}{1497--1502}.
%Type = Article
\bibitem[{Rohrer(2011)}]{Rohrer2011d}
\bibinfo{author}{G.~S. Rohrer},
\newblock \bibinfo{title}{{Grain boundary energy anisotropy: a review}},
\newblock \bibinfo{journal}{Journal of Materials Science} \bibinfo{volume}{46}
  (\bibinfo{year}{2011}) \bibinfo{pages}{5881--5895}.
%Type = Article
\bibitem[{Watanabe(2011)}]{Watanabe2011a}
\bibinfo{author}{T.~Watanabe},
\newblock \bibinfo{title}{{Grain boundary engineering: Historical perspective
  and future prospects}},
\newblock \bibinfo{journal}{Journal of Materials Science} \bibinfo{volume}{46}
  (\bibinfo{year}{2011}) \bibinfo{pages}{4095--4115}.
%Type = Article
\bibitem[{Hu et~al.(2018)Hu, Huang, Sumpter, Meletis, and Dumitrică}]{Hu2018b}
\bibinfo{author}{C.~Hu}, \bibinfo{author}{J.~Huang}, \bibinfo{author}{B.~G.
  Sumpter}, \bibinfo{author}{E.~Meletis}, \bibinfo{author}{T.~Dumitrică},
\newblock \bibinfo{title}{{Ab Initio Predictions of Strong Interfaces in
  Transition-Metal Carbides and Nitrides for Superhard Nanocomposite Coating
  Applications}},
\newblock \bibinfo{journal}{ACS Applied Nano Materials} \bibinfo{volume}{1}
  (\bibinfo{year}{2018}) \bibinfo{pages}{2029--2035}.
%Type = Article
\bibitem[{Watanabe and Tsurekawa(1999)}]{Watanabe1999a}
\bibinfo{author}{T.~Watanabe}, \bibinfo{author}{S.~Tsurekawa},
\newblock \bibinfo{title}{{The control of brittleness and development of
  desirable mechanical properties in polycrystalline systems by grain boundary
  engineering}},
\newblock \bibinfo{journal}{Acta Materialia} \bibinfo{volume}{47}
  (\bibinfo{year}{1999}) \bibinfo{pages}{4171--4185}.
%Type = Article
\bibitem[{Zheng et~al.(2018)Zheng, Tran, Li, Radhakrishnan, and
  Ong}]{Zheng2018}
\bibinfo{author}{H.~Zheng}, \bibinfo{author}{R.~Tran}, \bibinfo{author}{X.~G.
  Li}, \bibinfo{author}{B.~Radhakrishnan}, \bibinfo{author}{S.~P. Ong},
\newblock \bibinfo{title}{{Role of Zr in strengthening MoSi2 from density
  functional theory calculations}},
\newblock \bibinfo{journal}{Acta Materialia} \bibinfo{volume}{145}
  (\bibinfo{year}{2018}) \bibinfo{pages}{470--476}.
%Type = Article
\bibitem[{Lehockey and Palumbo(1997)}]{LEHOCKEY1997}
\bibinfo{author}{E.~M. Lehockey}, \bibinfo{author}{G.~Palumbo},
\newblock \bibinfo{title}{{On the creep behaviour of grain boundary engineered
  nickel 1}},
\newblock \bibinfo{journal}{Materials Science and Engineering: A}
  \bibinfo{volume}{237} (\bibinfo{year}{1997}) \bibinfo{pages}{168--172}.
%Type = Article
\bibitem[{Shi et~al.(2017)Shi, Hu, Zhang, Yuan, and Li}]{Shi2017}
\bibinfo{author}{P.~Shi}, \bibinfo{author}{R.~Hu}, \bibinfo{author}{T.~Zhang},
  \bibinfo{author}{L.~Yuan}, \bibinfo{author}{J.~Li},
\newblock \bibinfo{title}{{Grain boundary character distribution and its effect
  on corrosion of Ni–23Cr–16Mo superalloy}},
\newblock \bibinfo{journal}{Materials Science and Technology}
  \bibinfo{volume}{33} (\bibinfo{year}{2017}) \bibinfo{pages}{84--91}.
%Type = Article
\bibitem[{Kobayashi et~al.(2008)Kobayashi, Inomata, Kobayashi, Tsurekawa, and
  Watanabe}]{Kobayashi2008}
\bibinfo{author}{S.~Kobayashi}, \bibinfo{author}{T.~Inomata},
  \bibinfo{author}{H.~Kobayashi}, \bibinfo{author}{S.~Tsurekawa},
  \bibinfo{author}{T.~Watanabe},
\newblock \bibinfo{title}{{Effects of grain boundary- and triple
  junction-character on intergranular fatigue crack nucleation in
  polycrystalline aluminum}},
\newblock \bibinfo{journal}{Journal of Materials Science} \bibinfo{volume}{43}
  (\bibinfo{year}{2008}) \bibinfo{pages}{3792--3799}.
%Type = Article
\bibitem[{Lehockey et~al.(1998)Lehockey, Palumbo, and Lin}]{Lehockey1998}
\bibinfo{author}{E.~M. Lehockey}, \bibinfo{author}{G.~Palumbo},
  \bibinfo{author}{P.~Lin},
\newblock \bibinfo{title}{{Improving the weldability and service performance of
  nickel-and iron-based superalloys by grain boundary engineering}},
\newblock \bibinfo{journal}{Metallurgical and Materials Transactions A}
  \bibinfo{volume}{29} (\bibinfo{year}{1998}) \bibinfo{pages}{3069--3079}.
%Type = Article
\bibitem[{Was(1990)}]{Was1990}
\bibinfo{author}{G.~S. Was},
\newblock \bibinfo{title}{{Grain-Boundary Chemistry and Intergranular Fracture
  in Austenitic Nickel-Base Alloys—A Review}},
\newblock \bibinfo{journal}{CORROSION} \bibinfo{volume}{46}
  (\bibinfo{year}{1990}) \bibinfo{pages}{319--330}.
%Type = Article
\bibitem[{Watanabe(2011)}]{Watanabe2011}
\bibinfo{author}{T.~Watanabe},
\newblock \bibinfo{title}{{Grain boundary engineering: historical perspective
  and future prospects}},
\newblock \bibinfo{journal}{Journal of Materials Science} \bibinfo{volume}{46}
  (\bibinfo{year}{2011}) \bibinfo{pages}{4095--4115}.
%Type = Article
\bibitem[{Krupp et~al.(2003)Krupp, Kane, Liu, Dueber, Laird, and
  McMahon}]{KRUPP2003213}
\bibinfo{author}{U.~Krupp}, \bibinfo{author}{W.~M. Kane},
  \bibinfo{author}{X.~Liu}, \bibinfo{author}{O.~Dueber},
  \bibinfo{author}{C.~Laird}, \bibinfo{author}{C.~J. McMahon},
\newblock \bibinfo{title}{{The effect of grain-boundary-engineering-type
  processing on oxygen-induced cracking of IN718}},
\newblock \bibinfo{journal}{Materials Science and Engineering: A}
  \bibinfo{volume}{349} (\bibinfo{year}{2003}) \bibinfo{pages}{213--217}.
%Type = Article
\bibitem[{Pineau(2015)}]{Pineau2015}
\bibinfo{author}{A.~Pineau},
\newblock \bibinfo{title}{{Crossing grain boundaries in metals by slip bands,
  cleavage and fatigue cracks}},
\newblock \bibinfo{journal}{Philosophical Transactions of the Royal Society A:
  Mathematical, Physical and Engineering Sciences} \bibinfo{volume}{373}
  (\bibinfo{year}{2015}) \bibinfo{pages}{20140131}.
%Type = Article
\bibitem[{Rohrer et~al.(2010{\natexlab{a}})Rohrer, Li, Lee, Rollett, Groeber,
  and Uchic}]{Rohrer2010a}
\bibinfo{author}{G.~S. Rohrer}, \bibinfo{author}{J.~Li},
  \bibinfo{author}{S.~Lee}, \bibinfo{author}{A.~D. Rollett},
  \bibinfo{author}{M.~Groeber}, \bibinfo{author}{M.~D. Uchic},
\newblock \bibinfo{title}{{Deriving grain boundary character distributions and
  relative grain boundary energies from three-dimensional EBSD data}},
\newblock \bibinfo{journal}{Materials Science and Technology}
  \bibinfo{volume}{26} (\bibinfo{year}{2010}{\natexlab{a}})
  \bibinfo{pages}{661--669}.
%Type = Article
\bibitem[{Rohrer et~al.(2010{\natexlab{b}})Rohrer, Holm, Rollett, Foiles, Li,
  and Olmsted}]{Rohrer2010}
\bibinfo{author}{G.~S. Rohrer}, \bibinfo{author}{E.~A. Holm},
  \bibinfo{author}{A.~D. Rollett}, \bibinfo{author}{S.~M. Foiles},
  \bibinfo{author}{J.~Li}, \bibinfo{author}{D.~L. Olmsted},
\newblock \bibinfo{title}{{Comparing calculated and measured grain boundary
  energies in nickel}},
\newblock \bibinfo{journal}{Acta Materialia} \bibinfo{volume}{58}
  (\bibinfo{year}{2010}{\natexlab{b}}) \bibinfo{pages}{5063--5069}.
%Type = Article
\bibitem[{Hasson et~al.(1972)Hasson, Boos, Herbeuval, Biscondi, and
  Goux}]{HASSON1972115}
\bibinfo{author}{G.~Hasson}, \bibinfo{author}{J.-Y. Boos},
  \bibinfo{author}{I.~Herbeuval}, \bibinfo{author}{M.~Biscondi},
  \bibinfo{author}{C.~Goux},
\newblock \bibinfo{title}{{Theoretical and experimental determinations of grain
  boundary structures and energies: Correlation with various experimental
  results}},
\newblock \bibinfo{journal}{Surface Science} \bibinfo{volume}{31}
  (\bibinfo{year}{1972}) \bibinfo{pages}{115--137}.
%Type = Article
\bibitem[{Barmak et~al.(2006)Barmak, Kim, Kim, Archibald, Rohrer, Rollett,
  Kinderlehrer, Ta'asan, Zhang, and Srolovitz}]{BARMAK20061059}
\bibinfo{author}{K.~Barmak}, \bibinfo{author}{J.~Kim}, \bibinfo{author}{C.-S.
  Kim}, \bibinfo{author}{W.~E. Archibald}, \bibinfo{author}{G.~S. Rohrer},
  \bibinfo{author}{A.~D. Rollett}, \bibinfo{author}{D.~Kinderlehrer},
  \bibinfo{author}{S.~Ta'asan}, \bibinfo{author}{H.~Zhang},
  \bibinfo{author}{D.~J. Srolovitz},
\newblock \bibinfo{title}{{Grain boundary energy and grain growth in Al films:
  Comparison of experiments and simulations}},
\newblock \bibinfo{journal}{Scripta Materialia} \bibinfo{volume}{54}
  (\bibinfo{year}{2006}) \bibinfo{pages}{1059--1063}.
%Type = Article
\bibitem[{Gjostein and Rhines(1959)}]{GJOSTEIN1959319}
\bibinfo{author}{N.~A. Gjostein}, \bibinfo{author}{F.~N. Rhines},
\newblock \bibinfo{title}{{Absolute interfacial energies of [001] tilt and
  twist grain boundaries in copper}},
\newblock \bibinfo{journal}{Acta Metallurgica} \bibinfo{volume}{7}
  (\bibinfo{year}{1959}) \bibinfo{pages}{319--330}.
%Type = Article
\bibitem[{McLean(1973)}]{McLean1973}
\bibinfo{author}{M.~McLean},
\newblock \bibinfo{title}{{Grain-boundary energy of copper at 1030˚C}},
\newblock \bibinfo{journal}{Journal of Materials Science} \bibinfo{volume}{8}
  (\bibinfo{year}{1973}) \bibinfo{pages}{571--576}.
%Type = Article
\bibitem[{Chan and Balluffi(1986)}]{Chan1986}
\bibinfo{author}{S.~W. Chan}, \bibinfo{author}{R.~W. Balluffi},
\newblock \bibinfo{title}{{Study of energy vs misorientation for grain
  boundaries in gold by crystallite rotation method-II. Tilt boundaries and
  mixed boundaries}},
\newblock \bibinfo{journal}{Acta Metallurgica} \bibinfo{volume}{34}
  (\bibinfo{year}{1986}) \bibinfo{pages}{2191--2199}.
%Type = Article
\bibitem[{Miura et~al.(1994)Miura, Kato, and Mori}]{Miura1994}
\bibinfo{author}{H.~Miura}, \bibinfo{author}{M.~Kato},
  \bibinfo{author}{T.~Mori},
\newblock \bibinfo{title}{{Temperature dependence of the energy of Cu [110]
  symmetrical tilt grain boundaries}},
\newblock \bibinfo{journal}{Journal of Materials Science Letters}
  \bibinfo{volume}{13} (\bibinfo{year}{1994}) \bibinfo{pages}{46--48}.
%Type = Article
\bibitem[{Skidmore et~al.(2004)Skidmore, Buchheit, and Juhas}]{Skidmore2004a}
\bibinfo{author}{T.~Skidmore}, \bibinfo{author}{R.~G. Buchheit},
  \bibinfo{author}{M.~C. Juhas},
\newblock \bibinfo{title}{{Grain boundary energy vs. misorientation in
  Inconel{\textregistered} 600 alloy as measured by thermal groove and OIM
  analysis correlation}},
\newblock \bibinfo{journal}{Scripta Materialia} \bibinfo{volume}{50}
  (\bibinfo{year}{2004}) \bibinfo{pages}{873--877}.
%Type = Article
\bibitem[{Rohrer(2011)}]{Rohrer2011e}
\bibinfo{author}{G.~S. Rohrer},
\newblock \bibinfo{title}{{Measuring and interpreting the structure of
  grain-boundary networks}},
\newblock \bibinfo{journal}{Journal of the American Ceramic Society}
  \bibinfo{volume}{94} (\bibinfo{year}{2011}) \bibinfo{pages}{633--646}.
%Type = Article
\bibitem[{Amouyal et~al.(2005)Amouyal, Rabkin, and Mishin}]{Amouyal2005a}
\bibinfo{author}{Y.~Amouyal}, \bibinfo{author}{E.~Rabkin},
  \bibinfo{author}{Y.~Mishin},
\newblock \bibinfo{title}{{Correlation between grain boundary energy and
  geometry in Ni-rich NiAl}},
\newblock \bibinfo{journal}{Acta Materialia} \bibinfo{volume}{53}
  (\bibinfo{year}{2005}) \bibinfo{pages}{3795--3805}.
%Type = Article
\bibitem[{Li et~al.(2009)Li, Dillon, and Rohrer}]{Li2009}
\bibinfo{author}{J.~Li}, \bibinfo{author}{S.~J. Dillon}, \bibinfo{author}{G.~S.
  Rohrer},
\newblock \bibinfo{title}{{Relative grain boundary area and energy
  distributions in nickel}},
\newblock \bibinfo{journal}{Acta Materialia} \bibinfo{volume}{57}
  (\bibinfo{year}{2009}) \bibinfo{pages}{4304--4311}.
%Type = Article
\bibitem[{Liu et~al.(2013)Liu, Choi, Beladi, Nuhfer, Rohrer, and
  Barmak}]{Liu2013a}
\bibinfo{author}{X.~Liu}, \bibinfo{author}{D.~Choi},
  \bibinfo{author}{H.~Beladi}, \bibinfo{author}{N.~T. Nuhfer},
  \bibinfo{author}{G.~S. Rohrer}, \bibinfo{author}{K.~Barmak},
\newblock \bibinfo{title}{{The five-parameter grain boundary character
  distribution of nanocrystalline tungsten}},
\newblock \bibinfo{journal}{Scripta Materialia} \bibinfo{volume}{69}
  (\bibinfo{year}{2013}) \bibinfo{pages}{413--416}.
%Type = Article
\bibitem[{Beladi and Rohrer(2013)}]{BELADI20131404}
\bibinfo{author}{H.~Beladi}, \bibinfo{author}{G.~S. Rohrer},
\newblock \bibinfo{title}{{The relative grain boundary area and energy
  distributions in a ferritic steel determined from three-dimensional electron
  backscatter diffraction maps}},
\newblock \bibinfo{journal}{Acta Materialia} \bibinfo{volume}{61}
  (\bibinfo{year}{2013}) \bibinfo{pages}{1404--1412}.
%Type = Article
\bibitem[{Beladi et~al.(2014)Beladi, Nuhfer, and Rohrer}]{BELADI2014281}
\bibinfo{author}{H.~Beladi}, \bibinfo{author}{N.~T. Nuhfer},
  \bibinfo{author}{G.~S. Rohrer},
\newblock \bibinfo{title}{{The five-parameter grain boundary character and
  energy distributions of a fully austenitic high-manganese steel using three
  dimensional data}},
\newblock \bibinfo{journal}{Acta Materialia} \bibinfo{volume}{70}
  (\bibinfo{year}{2014}) \bibinfo{pages}{281--289}.
%Type = Article
\bibitem[{Kelly et~al.(2016)Kelly, Glowinski, Nuhfer, and Rohrer}]{Kelly2016}
\bibinfo{author}{M.~N. Kelly}, \bibinfo{author}{K.~Glowinski},
  \bibinfo{author}{N.~T. Nuhfer}, \bibinfo{author}{G.~S. Rohrer},
\newblock \bibinfo{title}{{The five parameter grain boundary character
  distribution of $\alpha$-Ti determined from three-dimensional orientation
  data}},
\newblock \bibinfo{journal}{Acta Materialia} \bibinfo{volume}{111}
  (\bibinfo{year}{2016}) \bibinfo{pages}{22--30}.
%Type = Article
\bibitem[{Wolf(1989)}]{Wolf1989e}
\bibinfo{author}{D.~Wolf},
\newblock \bibinfo{title}{{Structure-energy correlation for grain boundaries in
  F.C.C. metals—I. Boundaries on the (111) and (100) planes}},
\newblock \bibinfo{journal}{Acta Metallurgica} \bibinfo{volume}{37}
  (\bibinfo{year}{1989}) \bibinfo{pages}{1983--1993}.
%Type = Article
\bibitem[{Wolf and Phillpot(1989)}]{Wolf1989c}
\bibinfo{author}{D.~Wolf}, \bibinfo{author}{S.~Phillpot},
\newblock \bibinfo{title}{{Role of the densest lattice planes in the stability
  of crystalline interfaces: A computer simulation study}},
\newblock \bibinfo{journal}{Materials Science and Engineering: A}
  \bibinfo{volume}{A107} (\bibinfo{year}{1989}) \bibinfo{pages}{3--14}.
%Type = Article
\bibitem[{Wolf(1990{\natexlab{a}})}]{Wolf1989b}
\bibinfo{author}{D.~Wolf},
\newblock \bibinfo{title}{{Correlation between the energy and structure of
  grain boundaries in b.c.c. metals. II. Symmetrical tilt boundaries}},
\newblock \bibinfo{journal}{Philosophical Magazine A} \bibinfo{volume}{62}
  (\bibinfo{year}{1990}{\natexlab{a}}) \bibinfo{pages}{447--464}.
%Type = Article
\bibitem[{Wolf(1990{\natexlab{b}})}]{Wolf1990d}
\bibinfo{author}{D.~Wolf},
\newblock \bibinfo{title}{{A broken-bond model for grain boundaries in
  face-centered cubic metals}},
\newblock \bibinfo{journal}{Journal of Applied Physics} \bibinfo{volume}{68}
  (\bibinfo{year}{1990}{\natexlab{b}}) \bibinfo{pages}{3221--3236}.
%Type = Article
\bibitem[{Wolf(1990{\natexlab{c}})}]{Wolf1990e}
\bibinfo{author}{D.~Wolf},
\newblock \bibinfo{title}{{Structure-energy correlation for grain boundaries in
  F.C.C. metals-III. Symmetrical tilt boundaries}},
\newblock \bibinfo{journal}{Acta Metallurgica Et Materialia}
  \bibinfo{volume}{38} (\bibinfo{year}{1990}{\natexlab{c}})
  \bibinfo{pages}{781--790}.
%Type = Article
\bibitem[{Olmsted et~al.(2009)Olmsted, Foiles, and Holm}]{Olmsted2009}
\bibinfo{author}{D.~L. Olmsted}, \bibinfo{author}{S.~M. Foiles},
  \bibinfo{author}{E.~A. Holm},
\newblock \bibinfo{title}{{Survey of computed grain boundary properties in
  face-centered cubic metals: I. Grain boundary energy}},
\newblock \bibinfo{journal}{Acta Materialia} \bibinfo{volume}{57}
  (\bibinfo{year}{2009}) \bibinfo{pages}{3694--3703}.
%Type = Article
\bibitem[{Holm et~al.(2010)Holm, Olmsted, and Foiles}]{Holm2010a}
\bibinfo{author}{E.~A. Holm}, \bibinfo{author}{D.~L. Olmsted},
  \bibinfo{author}{S.~M. Foiles},
\newblock \bibinfo{title}{{Comparing grain boundary energies in face-centered
  cubic metals: Al, Au, Cu and Ni}},
\newblock \bibinfo{journal}{Scripta Materialia} \bibinfo{volume}{63}
  (\bibinfo{year}{2010}) \bibinfo{pages}{905--908}.
%Type = Article
\bibitem[{Tschopp et~al.(2015)Tschopp, Coleman, and McDowell}]{Tschopp2015}
\bibinfo{author}{M.~A. Tschopp}, \bibinfo{author}{S.~P. Coleman},
  \bibinfo{author}{D.~L. McDowell},
\newblock \bibinfo{title}{{Symmetric and asymmetric tilt grain boundary
  structure and energy in Cu and Al (and transferability to other fcc
  metals)}},
\newblock \bibinfo{journal}{Integrating Materials and Manufacturing Innovation}
  \bibinfo{volume}{4} (\bibinfo{year}{2015}) \bibinfo{pages}{11}.
%Type = Article
\bibitem[{Ratanaphan et~al.(2015)Ratanaphan, Olmsted, Bulatov, Holm, Rollett,
  and Rohrer}]{Ratanaphan2015b}
\bibinfo{author}{S.~Ratanaphan}, \bibinfo{author}{D.~L. Olmsted},
  \bibinfo{author}{V.~V. Bulatov}, \bibinfo{author}{E.~A. Holm},
  \bibinfo{author}{A.~D. Rollett}, \bibinfo{author}{G.~S. Rohrer},
\newblock \bibinfo{title}{{Grain boundary energies in body-centered cubic
  metals}},
\newblock \bibinfo{journal}{Acta Materialia} \bibinfo{volume}{88}
  (\bibinfo{year}{2015}) \bibinfo{pages}{346--354}.
%Type = Article
\bibitem[{Holm et~al.(2011)Holm, Rohrer, Foiles, Rollett, Miller, and
  Olmsted}]{Holm2011a}
\bibinfo{author}{E.~A. Holm}, \bibinfo{author}{G.~S. Rohrer},
  \bibinfo{author}{S.~M. Foiles}, \bibinfo{author}{A.~D. Rollett},
  \bibinfo{author}{H.~M. Miller}, \bibinfo{author}{D.~L. Olmsted},
\newblock \bibinfo{title}{{Validating computed grain boundary energies in fcc
  metals using the grain boundary character distribution}},
\newblock \bibinfo{journal}{Acta Materialia} \bibinfo{volume}{59}
  (\bibinfo{year}{2011}) \bibinfo{pages}{5250--5256}.
%Type = Article
\bibitem[{Scheiber et~al.(2016)Scheiber, Pippan, Puschnig, and
  Romaner}]{Scheiber2016f}
\bibinfo{author}{D.~Scheiber}, \bibinfo{author}{R.~Pippan},
  \bibinfo{author}{P.~Puschnig}, \bibinfo{author}{L.~Romaner},
\newblock \bibinfo{title}{{Ab initio calculations of grain boundaries in bcc
  metals}},
\newblock \bibinfo{journal}{Modelling and Simulation in Materials Science and
  Engineering} \bibinfo{volume}{24} (\bibinfo{year}{2016})
  \bibinfo{pages}{035013}.
%Type = Article
\bibitem[{Wang et~al.(2018)Wang, Madsen, and Drautz}]{Wang2018i}
\bibinfo{author}{J.~Wang}, \bibinfo{author}{G.~K.~H. Madsen},
  \bibinfo{author}{R.~Drautz},
\newblock \bibinfo{title}{Grain boundaries in bcc-fe: a density-functional
  theory and tight-binding study},
\newblock \bibinfo{journal}{Modelling and Simulation in Materials Science and
  Engineering} \bibinfo{volume}{26} (\bibinfo{year}{2018})
  \bibinfo{pages}{025008}.
%Type = Article
\bibitem[{Bean and McKenna(2016)}]{Bean2016a}
\bibinfo{author}{J.~J. Bean}, \bibinfo{author}{K.~P. McKenna},
\newblock \bibinfo{title}{{Origin of differences in the excess volume of copper
  and nickel grain boundaries}},
\newblock \bibinfo{journal}{Acta Materialia} \bibinfo{volume}{110}
  (\bibinfo{year}{2016}) \bibinfo{pages}{246--257}.
%Type = Article
\bibitem[{Ong et~al.(2013)Ong, Richards, Jain, Hautier, Kocher, Cholia, Gunter,
  Chevrier, Persson, and Ceder}]{Ong2013}
\bibinfo{author}{S.~P. Ong}, \bibinfo{author}{W.~D. Richards},
  \bibinfo{author}{A.~Jain}, \bibinfo{author}{G.~Hautier},
  \bibinfo{author}{M.~Kocher}, \bibinfo{author}{S.~Cholia},
  \bibinfo{author}{D.~Gunter}, \bibinfo{author}{V.~L. Chevrier},
  \bibinfo{author}{K.~A. Persson}, \bibinfo{author}{G.~Ceder},
\newblock \bibinfo{title}{{Python Materials Genomics (pymatgen): A robust,
  open-source python library for materials analysis}},
\newblock \bibinfo{journal}{Computational Materials Science}
  \bibinfo{volume}{68} (\bibinfo{year}{2013}) \bibinfo{pages}{314--319}.
%Type = Article
\bibitem[{Ong et~al.(2015)Ong, Cholia, Jain, Brafman, Gunter, Ceder, and
  Persson}]{Ong2015g}
\bibinfo{author}{S.~P. Ong}, \bibinfo{author}{S.~Cholia},
  \bibinfo{author}{A.~Jain}, \bibinfo{author}{M.~Brafman},
  \bibinfo{author}{D.~Gunter}, \bibinfo{author}{G.~Ceder},
  \bibinfo{author}{K.~A. Persson},
\newblock \bibinfo{title}{{The Materials Application Programming Interface
  (API): A simple, flexible and efficient API for materials data based on
  REpresentational State Transfer (REST) principles}},
\newblock \bibinfo{journal}{Computational Materials Science}
  \bibinfo{volume}{97} (\bibinfo{year}{2015}) \bibinfo{pages}{209--215}.
%Type = Article
\bibitem[{Grimmer(1976)}]{Grimmer:a28679}
\bibinfo{author}{H.~Grimmer},
\newblock \bibinfo{title}{{Coincidence-site lattices}},
\newblock \bibinfo{journal}{Acta Crystallographica Section A}
  \bibinfo{volume}{32} (\bibinfo{year}{1976}) \bibinfo{pages}{783--785}.
%Type = Incollection
\bibitem[{Lej{\v{c}}ek(2010)}]{Lejcek2010a}
\bibinfo{author}{P.~Lej{\v{c}}ek},
\newblock \bibinfo{title}{{Grain Boundaries: Description, Structure and
  Thermodynamics}},
\newblock in: \bibinfo{booktitle}{Grain boundary Segregation in Metals},
  \bibinfo{year}{2010}, pp. \bibinfo{pages}{5--24}.
%Type = Article
\bibitem[{M{\"{o}}ller and Bitzek(2014)}]{MOLLER20141}
\bibinfo{author}{J.~J. M{\"{o}}ller}, \bibinfo{author}{E.~Bitzek},
\newblock \bibinfo{title}{{Fracture toughness and bond trapping of grain
  boundary cracks}},
\newblock \bibinfo{journal}{Acta Materialia} \bibinfo{volume}{73}
  (\bibinfo{year}{2014}) \bibinfo{pages}{1--11}.
%Type = Article
\bibitem[{Coffman and Sethna(2008)}]{Coffman2008b}
\bibinfo{author}{V.~R. Coffman}, \bibinfo{author}{J.~P. Sethna},
\newblock \bibinfo{title}{{Grain boundary energies and cohesive strength as a
  function of geometry}},
\newblock \bibinfo{journal}{Physical Review B} \bibinfo{volume}{77}
  (\bibinfo{year}{2008}) \bibinfo{pages}{144111}.
%Type = Article
\bibitem[{Grujicic et~al.(1997)Grujicic, Zhao, and Krasko}]{GRUJICIC1997341}
\bibinfo{author}{M.~Grujicic}, \bibinfo{author}{H.~Zhao},
  \bibinfo{author}{G.~L. Krasko},
\newblock \bibinfo{title}{{Atomistic simulation of ∑3 (111) grain boundary
  fracture in tungsten containing various impurities}},
\newblock \bibinfo{journal}{International Journal of Refractory Metals and Hard
  Materials} \bibinfo{volume}{15} (\bibinfo{year}{1997})
  \bibinfo{pages}{341--355}.
%Type = Article
\bibitem[{Gumbsch(1999)}]{GUMBSCH199972}
\bibinfo{author}{P.~Gumbsch},
\newblock \bibinfo{title}{{Atomistic modelling of diffusion-controlled
  interfacial decohesion}},
\newblock \bibinfo{journal}{Materials Science and Engineering: A}
  \bibinfo{volume}{260} (\bibinfo{year}{1999}) \bibinfo{pages}{72--79}.
%Type = Article
\bibitem[{Tran et~al.(2016)Tran, Xu, Radhakrishnan, Winston, Sun, Persson, and
  Ong}]{Tran2016a}
\bibinfo{author}{R.~Tran}, \bibinfo{author}{Z.~Xu},
  \bibinfo{author}{B.~Radhakrishnan}, \bibinfo{author}{D.~Winston},
  \bibinfo{author}{W.~Sun}, \bibinfo{author}{K.~A. Persson},
  \bibinfo{author}{S.~P. Ong},
\newblock \bibinfo{title}{{Data Descripter: Surface energies of elemental
  crystals}},
\newblock \bibinfo{journal}{Scientific Data} \bibinfo{volume}{3}
  (\bibinfo{year}{2016}) \bibinfo{pages}{1--13}.
%Type = Article
\bibitem[{Kohn and Sham(1965)}]{Kohn1965}
\bibinfo{author}{W.~Kohn}, \bibinfo{author}{L.~J. Sham},
\newblock \bibinfo{title}{Self-consistent equations including exchange and
  correlation effects},
\newblock \bibinfo{journal}{Phys. Rev.} \bibinfo{volume}{140}
  (\bibinfo{year}{1965}) \bibinfo{pages}{A1133--A1138}.
%Type = Article
\bibitem[{Kresse and Furthm{\"{u}}ller(1996)}]{Kresse1996a}
\bibinfo{author}{G.~Kresse}, \bibinfo{author}{J.~Furthm{\"{u}}ller},
\newblock \bibinfo{title}{{Efficient iterative schemes for ab initio
  total-energy calculations using a plane-wave basis set}},
\newblock \bibinfo{journal}{Physical Review B} \bibinfo{volume}{54}
  (\bibinfo{year}{1996}) \bibinfo{pages}{11169--11186}.
%Type = Article
\bibitem[{Bl{\"{o}}chl(1994)}]{Blochl1994}
\bibinfo{author}{P.~E. Bl{\"{o}}chl},
\newblock \bibinfo{title}{{Projector augmented-wave method}},
\newblock \bibinfo{journal}{Physical Review B} \bibinfo{volume}{50}
  (\bibinfo{year}{1994}) \bibinfo{pages}{17953--17979}.
%Type = Article
\bibitem[{Perdew et~al.(1996)Perdew, Burke, and Ernzerhof}]{Perdew1996}
\bibinfo{author}{J.~P. Perdew}, \bibinfo{author}{K.~Burke},
  \bibinfo{author}{M.~Ernzerhof},
\newblock \bibinfo{title}{{Generalized Gradient Approximation Made Simple}},
\newblock \bibinfo{journal}{Physical Review Letters} \bibinfo{volume}{77}
  (\bibinfo{year}{1996}) \bibinfo{pages}{3865--3868}.
%Type = Article
\bibitem[{Mathew et~al.(2017)Mathew, Montoya, Faghaninia, Dwarakanath, Aykol,
  Tang, Chu, Smidt, Bocklund, Horton, Dagdelen, Wood, Liu, Neaton, Ping,
  Persson, Jain, Ong, Persson, Jain, Ping, Persson, and Jain}]{Mathew2017}
\bibinfo{author}{K.~Mathew}, \bibinfo{author}{J.~H. Montoya},
  \bibinfo{author}{A.~Faghaninia}, \bibinfo{author}{S.~Dwarakanath},
  \bibinfo{author}{M.~Aykol}, \bibinfo{author}{H.~Tang}, \bibinfo{author}{I.-h.
  Chu}, \bibinfo{author}{T.~Smidt}, \bibinfo{author}{B.~Bocklund},
  \bibinfo{author}{M.~Horton}, \bibinfo{author}{J.~Dagdelen},
  \bibinfo{author}{B.~Wood}, \bibinfo{author}{Z.-k. Liu},
  \bibinfo{author}{J.~Neaton}, \bibinfo{author}{S.~Ping},
  \bibinfo{author}{K.~Persson}, \bibinfo{author}{A.~Jain},
  \bibinfo{author}{S.~P. Ong}, \bibinfo{author}{K.~Persson},
  \bibinfo{author}{A.~Jain}, \bibinfo{author}{S.~Ping},
  \bibinfo{author}{K.~Persson}, \bibinfo{author}{A.~Jain},
\newblock \bibinfo{title}{{Atomate: A high-level interface to generate,
  execute, and analyze computational materials science workflows}},
\newblock \bibinfo{journal}{Computational Materials Science}
  \bibinfo{volume}{139} (\bibinfo{year}{2017}) \bibinfo{pages}{140--152}.
%Type = Article
\bibitem[{Jain et~al.(2015)Jain, Ong, Chen, Medasani, Qu, Kocher, Brafman,
  Petretto, Rignanese, Hautier, Gunter, and Persson}]{Jain2015b}
\bibinfo{author}{A.~Jain}, \bibinfo{author}{S.~P. Ong},
  \bibinfo{author}{W.~Chen}, \bibinfo{author}{B.~Medasani},
  \bibinfo{author}{X.~Qu}, \bibinfo{author}{M.~Kocher},
  \bibinfo{author}{M.~Brafman}, \bibinfo{author}{G.~Petretto},
  \bibinfo{author}{G.-M. Rignanese}, \bibinfo{author}{G.~Hautier},
  \bibinfo{author}{D.~Gunter}, \bibinfo{author}{K.~A. Persson},
\newblock \bibinfo{title}{{FireWorks: a dynamic workflow system designed for
  high-throughput applications}},
\newblock \bibinfo{journal}{Concurrency and Computation: Practice and
  Experience} \bibinfo{volume}{27} (\bibinfo{year}{2015})
  \bibinfo{pages}{5037--5059}.
%Type = Article
\bibitem[{Wolf(1991)}]{Wolf1991}
\bibinfo{author}{D.~Wolf},
\newblock \bibinfo{title}{{Structure and energy of general grain boundaries in
  bcc metals}},
\newblock \bibinfo{journal}{Journal of Applied Physics} \bibinfo{volume}{69}
  (\bibinfo{year}{1991}) \bibinfo{pages}{185--196}.
%Type = Article
\bibitem[{Saylor et~al.(2004)Saylor, {El Dasher}, Rollett, and
  Rohrer}]{Saylor2004a}
\bibinfo{author}{D.~M. Saylor}, \bibinfo{author}{B.~S. {El Dasher}},
  \bibinfo{author}{A.~D. Rollett}, \bibinfo{author}{G.~S. Rohrer},
\newblock \bibinfo{title}{{Distribution of grain boundaries in aluminum as a
  function of five macroscopic parameters}},
\newblock \bibinfo{journal}{Acta Materialia} \bibinfo{volume}{52}
  (\bibinfo{year}{2004}) \bibinfo{pages}{3649--3655}.
%Type = Article
\bibitem[{Beladi and Rohrer(2013)}]{Beladi2013a}
\bibinfo{author}{H.~Beladi}, \bibinfo{author}{G.~S. Rohrer},
\newblock \bibinfo{title}{{The distribution of grain boundary planes in
  interstitial free steel}},
\newblock \bibinfo{journal}{Metallurgical and Materials Transactions A:
  Physical Metallurgy and Materials Science} \bibinfo{volume}{44}
  (\bibinfo{year}{2013}) \bibinfo{pages}{115--124}.
%Type = Article
\bibitem[{Bhattacharya et~al.(2013)Bhattacharya, Tanaka, Shiihara, and
  Kohyama}]{Bhattacharya2013}
\bibinfo{author}{S.~K. Bhattacharya}, \bibinfo{author}{S.~Tanaka},
  \bibinfo{author}{Y.~Shiihara}, \bibinfo{author}{M.~Kohyama},
\newblock \bibinfo{title}{Ab initiostudy of symmetrical tilt grain boundaries
  in bcc fe: structural units, magnetic moments, interfacial bonding, local
  energy and local stress},
\newblock \bibinfo{journal}{Journal of Physics: Condensed Matter}
  \bibinfo{volume}{25} (\bibinfo{year}{2013}) \bibinfo{pages}{135004}.
%Type = Article
\bibitem[{Wachowicz et~al.(2010)Wachowicz, Ossowski, and
  Kiejna}]{Wachowicz2010}
\bibinfo{author}{E.~Wachowicz}, \bibinfo{author}{T.~Ossowski},
  \bibinfo{author}{A.~Kiejna},
\newblock \bibinfo{title}{{Cohesive and magnetic properties of grain boundaries
  in bcc Fe with Cr additions}},
\newblock \bibinfo{journal}{Phys. Rev. B} \bibinfo{volume}{81}
  (\bibinfo{year}{2010}) \bibinfo{pages}{94104}.
%Type = Article
\bibitem[{Gao et~al.(2009)Gao, Fu, Samaras, Sch{\"{a}}ublin, Victoria, and
  Hoffelner}]{Gao2009}
\bibinfo{author}{N.~Gao}, \bibinfo{author}{C.~C. Fu},
  \bibinfo{author}{M.~Samaras}, \bibinfo{author}{R.~Sch{\"{a}}ublin},
  \bibinfo{author}{M.~Victoria}, \bibinfo{author}{W.~Hoffelner},
\newblock \bibinfo{title}{{Multiscale modelling of bi-crystal grain boundaries
  in bcc iron}},
\newblock \bibinfo{journal}{Journal of Nuclear Materials} \bibinfo{volume}{385}
  (\bibinfo{year}{2009}) \bibinfo{pages}{262--267}.
%Type = Article
\bibitem[{Du et~al.(2011)Du, Ismer, Jutta, Hickel, Neugebauer, and
  Drautz}]{Du2011}
\bibinfo{author}{Y.~A. Du}, \bibinfo{author}{L.~Ismer},
  \bibinfo{author}{R.~Jutta}, \bibinfo{author}{T.~Hickel},
  \bibinfo{author}{J.~Neugebauer}, \bibinfo{author}{R.~Drautz},
\newblock \bibinfo{title}{{First-principles study on the interaction of H
  interstitials with grain boundaries in $\alpha$ and $\gamma$-Fe}},
\newblock \bibinfo{journal}{Phys. Rev. B} \bibinfo{volume}{84}
  (\bibinfo{year}{2011}) \bibinfo{pages}{144121}.
%Type = Article
\bibitem[{{\v{C}}{\'{a}}k et~al.(2008){\v{C}}{\'{a}}k, {\v{S}}ob, and
  Hafner}]{Cak2008}
\bibinfo{author}{M.~{\v{C}}{\'{a}}k}, \bibinfo{author}{M.~{\v{S}}ob},
  \bibinfo{author}{J.~Hafner},
\newblock \bibinfo{title}{{First-principles study of magnetism at grain
  boundaries in iron and nickel}},
\newblock \bibinfo{journal}{Physical Review B - Condensed Matter and Materials
  Physics} \bibinfo{volume}{78} (\bibinfo{year}{2008}) \bibinfo{pages}{1--10}.
%Type = Article
\bibitem[{Ochs et~al.(2000)Ochs, Els{\"{a}}sser, Mrovec, Vitek, Belak, and
  Moriarty}]{Ochs2000a}
\bibinfo{author}{T.~Ochs}, \bibinfo{author}{C.~Els{\"{a}}sser},
  \bibinfo{author}{M.~Mrovec}, \bibinfo{author}{V.~Vitek},
  \bibinfo{author}{J.~Belak}, \bibinfo{author}{J.~A. Moriarty},
\newblock \bibinfo{title}{{Symmetrical tilt grain boundaries in bcc transition
  metals: Comparison of semiempirical with ab-initio total-energy
  calculations}},
\newblock \bibinfo{journal}{Philosophical Magazine A: Physics of Condensed
  Matter, Structure, Defects and Mechanical Properties} \bibinfo{volume}{80}
  (\bibinfo{year}{2000}) \bibinfo{pages}{2405--2423}.
%Type = Article
\bibitem[{Wright and Atlas(1994)}]{Wright1994}
\bibinfo{author}{A.~F. Wright}, \bibinfo{author}{S.~R. Atlas},
\newblock \bibinfo{title}{{Density-functional calculations for grain boundaries
  in aluminum}},
\newblock \bibinfo{journal}{Phys. Rev. B} \bibinfo{volume}{50}
  (\bibinfo{year}{1994}) \bibinfo{pages}{15248--15260}.
%Type = Article
\bibitem[{Plimpton(1995)}]{PLIMPTON19951}
\bibinfo{author}{S.~Plimpton},
\newblock \bibinfo{title}{{Fast Parallel Algorithms for Short-Range Molecular
  Dynamics}},
\newblock \bibinfo{journal}{Journal of Computational Physics}
  \bibinfo{volume}{117} (\bibinfo{year}{1995}) \bibinfo{pages}{1--19}.
%Type = Article
\bibitem[{Chen et~al.(2017)Chen, Deng, Tran, Tang, Chu, and Ong}]{Chen2017}
\bibinfo{author}{C.~Chen}, \bibinfo{author}{Z.~Deng},
  \bibinfo{author}{R.~Tran}, \bibinfo{author}{H.~Tang}, \bibinfo{author}{I.-h.
  Chu}, \bibinfo{author}{S.~P. Ong},
\newblock \bibinfo{title}{{Accurate force field for molybdenum by machine
  learning large materials data}},
\newblock \bibinfo{journal}{Physical Review Materials} \bibinfo{volume}{1}
  (\bibinfo{year}{2017}) \bibinfo{pages}{043603}.
%Type = Article
\bibitem[{Li et~al.(2018)Li, Hu, Chen, Deng, Luo, and Ong}]{Li2018j}
\bibinfo{author}{X.-G. Li}, \bibinfo{author}{C.~Hu}, \bibinfo{author}{C.~Chen},
  \bibinfo{author}{Z.~Deng}, \bibinfo{author}{J.~Luo}, \bibinfo{author}{S.~P.
  Ong},
\newblock \bibinfo{title}{{Quantum-accurate spectral neighbor analysis
  potential models for Ni-Mo binary alloys and fcc metals}},
\newblock \bibinfo{journal}{Phys. Rev. B} \bibinfo{volume}{98}
  (\bibinfo{year}{2018}) \bibinfo{pages}{94104}.
%Type = Misc
\bibitem[{MatWeb(2019)}]{web:exp_UTS}
\bibinfo{author}{MatWeb}, \bibinfo{title}{http://www.matweb.com/},
  \bibinfo{year}{2019}. \bibinfo{note}{Last accessed 14 April 2019}.
%Type = Article
\bibitem[{de~Jong et~al.(2015)de~Jong, Chen, Angsten, Jain, Notestine, Gamst,
  Sluiter, {Krishna Ande}, van~der Zwaag, Plata, Toher, Curtarolo, Ceder,
  Persson, and Asta}]{DeJong2015}
\bibinfo{author}{M.~de~Jong}, \bibinfo{author}{W.~Chen},
  \bibinfo{author}{T.~Angsten}, \bibinfo{author}{A.~Jain},
  \bibinfo{author}{R.~Notestine}, \bibinfo{author}{A.~Gamst},
  \bibinfo{author}{M.~Sluiter}, \bibinfo{author}{C.~{Krishna Ande}},
  \bibinfo{author}{S.~van~der Zwaag}, \bibinfo{author}{J.~J. Plata},
  \bibinfo{author}{C.~Toher}, \bibinfo{author}{S.~Curtarolo},
  \bibinfo{author}{G.~Ceder}, \bibinfo{author}{K.~A. Persson},
  \bibinfo{author}{M.~Asta},
\newblock \bibinfo{title}{{Charting the complete elastic properties of
  inorganic crystalline compounds}},
\newblock \bibinfo{journal}{Scientific Data} \bibinfo{volume}{2}
  (\bibinfo{year}{2015}) \bibinfo{pages}{150009}.
%Type = Article
\bibitem[{Read et~al.(1950)Read, Shockley, and Number}]{PhysRev.78.275}
\bibinfo{author}{W.~T. Read}, \bibinfo{author}{W.~Shockley},
  \bibinfo{author}{V.~Number},
\newblock \bibinfo{title}{{Dislocation Models of Crystal Grain Boundaries}},
\newblock \bibinfo{journal}{Phys. Rev.} \bibinfo{volume}{78}
  (\bibinfo{year}{1950}) \bibinfo{pages}{275--289}.

\end{thebibliography}






% \end{thebibliography}

\end{document}

\begin{frontmatter}


\title{Supplementary Information\\Grain Boundary Properties of Elemental Metals}



\author[UCSD]{Hui Zheng \fnref{cor2}}
\author[UCSD]{Xiang-Guo Li \fnref{cor2}}
\author[UCSD]{Richard Tran}
\author[UCSD]{Chi Chen}
\fntext[cor2]{These authors contributed equally}

\author[LBNL]{Matthew Horton}
\author[LBNL]{Donny Winston}
\author[LBNL,UCB]{Kristin Aslaug Persson}

\author[UCSD]{Shyue Ping Ong \corref{cor1}}
\ead{ongsp@eng.ucsd.edu}
\cortext[cor1]{Corresponding author}

\address[UCSD]{Department of NanoEngineering, University of California San Diego, 9500 Gilman Dr, Mail Code 0448, La Jolla, CA 92093-0448, United States}
\address[LBNL]{Environmental Energy Technologies Division, Lawrence Berkeley National Laboratory, Berkeley,  CA 94720, United States}
\address[UCB]{Department of Materials Science \& Engineering, University of California Berkeley, Berkeley,  CA 94720-1760, United States}
\end{frontmatter}
% \newpage

\begin{figure}[H]
\centering\includegraphics[width=0.85\linewidth]{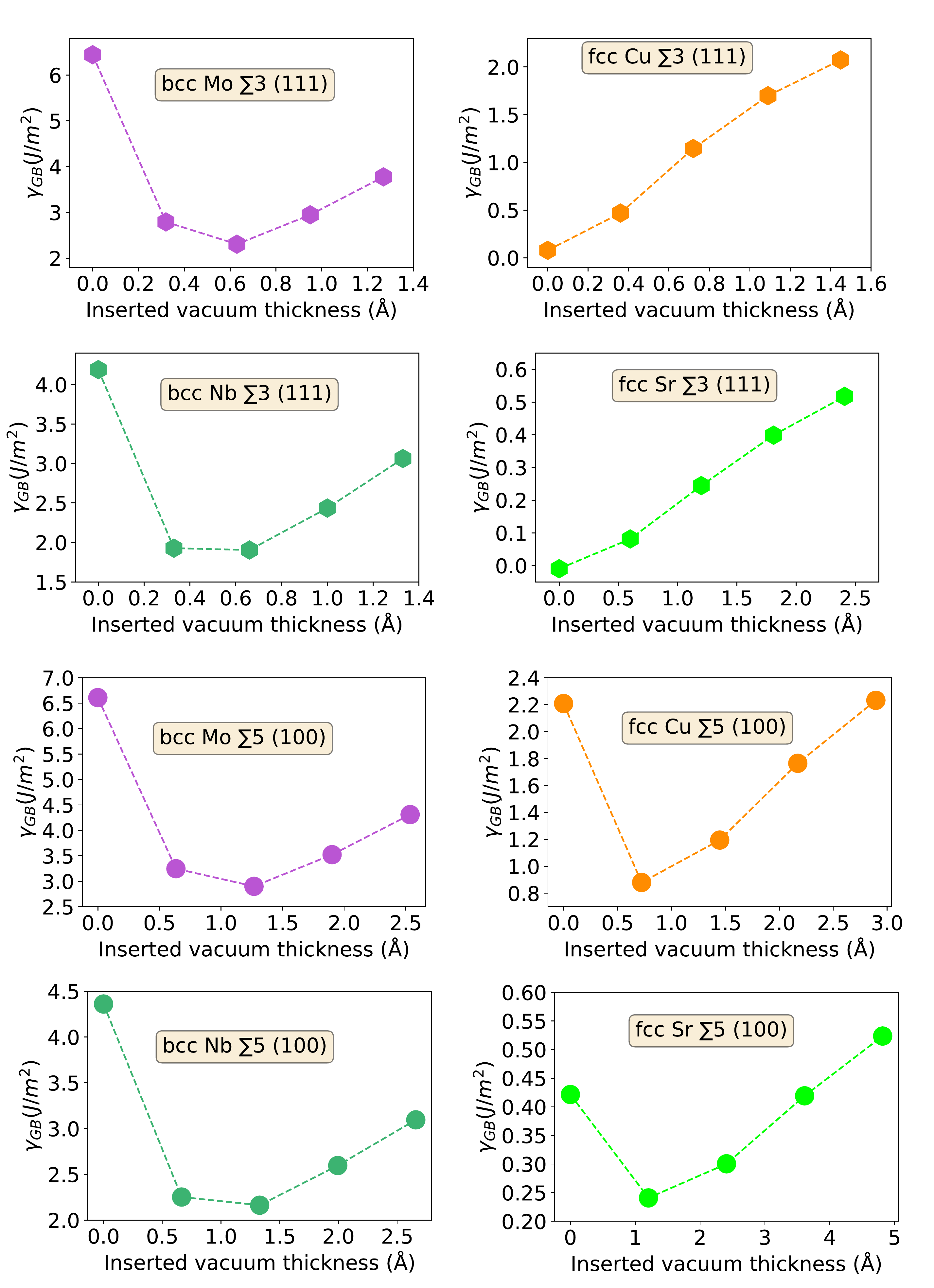}
\caption{\label{SI:c_shift_test} Lattice translation tests (along $c$ direction) for twist GBs. }
\end{figure}
\newpage
\begin{figure}[H]
\centering\includegraphics[width=0.85\linewidth]{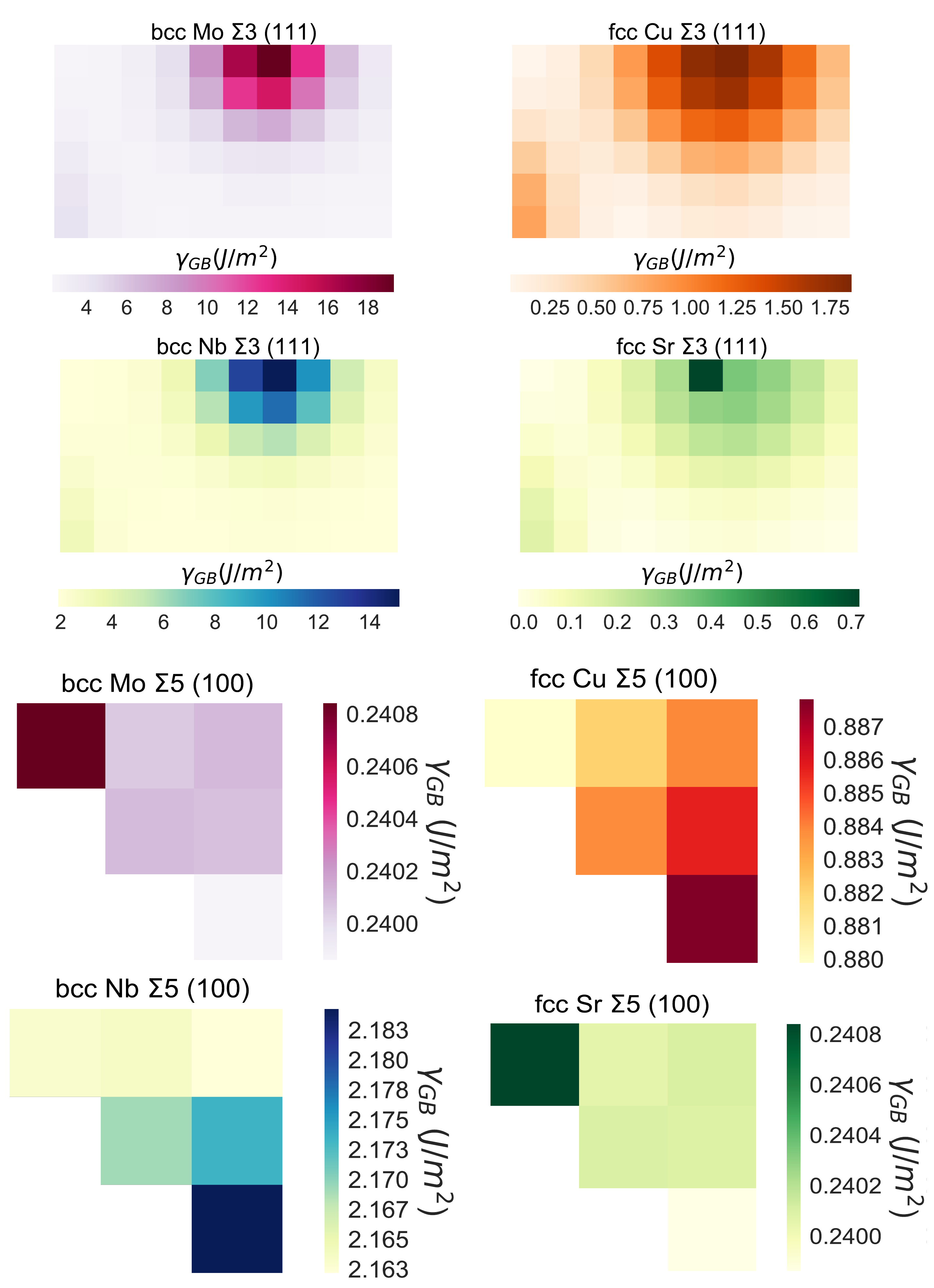}
\caption{\label{SI:ab_shift_test} Lattice translation tests (along $a$ and $b$ directions) for twist GBs. }
\end{figure}

\begin{center}
\begin{longtable}[c]{ccccccccccc}
\caption{Grain boundary energies of cubic elements}
\label{SI:gb_energies_cubic}\\
\hline
sigma &  $\Sigma$3 &  $\Sigma$3 &  $\Sigma$3&  $\Sigma$5&  $\Sigma$5 &  $\Sigma$5 &  $\Sigma$7 &  $\Sigma$7 &  $\Sigma$9 &  $\Sigma$9\\\hline  
plane &  (011) &  (111) &  (112) &  (001) &  (012) &  (013) &  (111) &  (123) & (011) & (122) \\\hline  
\endfirsthead
\hline
\endhead
\hline
\endfoot
\hline
\endlastfoot

Ba &    0.045 &    0.182 &    0.026 &    0.279 &    0.144 &    0.161 &    0.210 &    0.135 &    0.098 &    0.187  \\\hline 
Cr &    0.675 &    1.981 &    0.645 &    2.646 &    2.195 &    1.977 &    2.335 &    1.901 &    1.178 &    2.253  \\\hline 
Cs &   -0.002 &    0.013 &   -0.000 &    0.021 &    0.013 &    0.021 &    0.025 &    0.011 &   -0.004 &    0.019 \\\hline 
Fe &    0.508 &    1.598 &    0.423 &    2.243 &    1.892 &    1.560 &    1.904 &    1.419 &    0.970 &    1.754 \\\hline 
K  &    0.032 &    0.077 &    0.025 &    0.088 &    0.057 &    0.073 &    0.058 &    0.063 &    0.023 &    0.071 \\\hline 
Li &    0.046 &    0.171 &    0.054 &    0.269 &    0.150 &    0.156 &    0.187 &    0.142 &    0.082 &    0.202 \\\hline 
Mo &    0.503 &    1.743 &    0.480 &    2.432 &    2.029 &    1.727 &    2.210 &    1.732 &    1.040 &    2.064 \\\hline 
Na &    0.063 &    0.123 &    0.044 &    0.138 &    0.082 &    0.132 &    0.123 &    0.086 &    0.074 &    0.129 \\\hline 
Nb &    0.294 &    1.286 &    0.252 &    1.492 &    1.242 &    1.200 &    1.343 &    1.018 &    0.598 &    1.271 \\\hline 
Rb &    0.001 &    0.048 &    0.008 &    0.060 &    0.034 &    0.045 &    0.037 &    0.026 &    0.008 &    0.043 \\\hline 
Ta &    0.334 &    1.526 &    0.289 &    1.925 &    1.412 &    1.438 &    1.558 &    1.216 &    0.732 &    1.508 \\\hline 
V  &    0.322 &    1.176 &    0.258 &    1.383 &    1.204 &    1.262 &    1.255 &    0.941 &    0.611 &    1.210 \\\hline 
W  &    0.714 &    2.242 &    0.665 &    3.205 &    2.654 &    2.204 &    2.866 &    2.327 &    1.400 &    2.806 \\\hline 
Ag &    0.545 &    0.069 &    0.427 &    0.419 &    0.593 &    0.553 &    0.208 &    0.542 &    0.710 &    0.507 \\\hline 
Al &    0.456 &   -0.004 &    0.310 &    0.383 &    0.532 &    0.476 &    0.135 &    0.501 &    0.713 &    0.430 \\\hline 
Au &    0.458 &    0.026 &    0.338 &    0.323 &    0.520 &    0.447 &    0.175 &    0.441 &    0.607 &    0.393 \\\hline 
Ca &    0.239 &    0.018 &    0.204 &    0.262 &    0.303 &    0.302 &    0.109 &    0.307 &    0.383 &    0.291 \\\hline 
Ce &    0.496 &    0.219 &    0.489 &    0.578 &    0.516 &    0.580 &    0.285 &    0.539 &    0.622 &    0.485 \\\hline 
Cu &    0.848 &    0.071 &    0.634 &    0.751 &    0.997 &    0.882 &    0.370 &    0.916 &    1.166 &    0.856 \\\hline 
Ir &    1.857 &    0.352 &    1.616 &    1.296 &    2.186 &    1.728 &    0.856 &    1.866 &    2.246 &    1.578 \\\hline 
Ni &    1.210 &   -0.007 &    1.150 &    1.088 &    1.383 &    1.262 &    0.512 &    1.327 &    1.668 &    1.161 \\\hline 
Pb &    0.242 &    0.067 &    0.221 &    0.224 &    0.281 &    0.243 &    0.062 &    0.229 &    0.313 &    0.243 \\\hline 
Pd &    0.886 &    0.072 &    0.682 &    0.718 &    1.003 &    0.898 &    0.319 &    0.950 &    1.189 &    0.852 \\\hline 
Pt &    0.864 &    0.176 &    0.675 &    0.616 &    1.094 &    0.889 &    0.279 &    0.885 &    1.165 &    0.795 \\\hline 
Rh &    1.552 &    0.082 &    1.250 &    1.172 &    1.680 &    1.447 &    0.655 &    1.541 &    1.854 &    1.311 \\\hline 
Sr &    0.187 &   -0.010 &    0.159 &    0.199 &    0.234 &    0.237 &    0.088 &    0.240 &    0.310 &    0.225 \\\hline 
Th &    0.684 &    0.175 &    0.662 &    0.775 &    0.758 &    0.823 &    0.385 &    0.817 &    0.941 &    0.770 \\\hline 
Yb &    0.243 &    0.003 &    0.198 &    0.257 &    0.301 &    0.302 &    0.130 &    0.310 &    0.397 &    0.294 \\\hline 
\end{longtable}
\end{center}

\begin{center}
\begin{longtable}[c]{cccccccccc}
\caption{Grain boundary energies of hcp elements}
\label{SI:gb_energies_hcp}\\
\hline
{} &        Be &        Cd &        Co &       Dy &        Er &     Gd&   Hf &        Ho &        La \\\hline 
$\Sigma$7(0001)  &   1.121 &  0.116 &  0.722 &  0.264 &  0.318 & 0.201 & 0.411 &  0.297 &  0.139  \\\hline 
{}  & Lu &   Mg &        Nd &        Os &        Pm &        Pr &        Re &        Ru &        Sc \\\hline 
$\Sigma$7(0001) &  0.342 & 0.200 &  0.208 &  1.578 &  0.217 &  0.195 &  0.970 &  1.119 &  0.294  \\\hline 

{} &        Sm &     Tb &      Tc &        Ti &        Tl &        Tm &        Y &    Zn &   Zr  \\\hline 
$\Sigma$7(0001) &   0.227 &  0.242 &  0.671 &  0.365 &  0.074 &  0.328 &  0.216 &  0.141 &  0.314 {}\\\hline 
\end{longtable}
\end{center}

\begin{center}
\begin{longtable}[c]{ccccccccccc}
\caption{Work of separation ($W_{sep}$) of cubic elements}
\label{SI:Wsep_cubic}\\
\hline
sigma &  $\Sigma$3 &  $\Sigma$3 &  $\Sigma$3&  $\Sigma$5&  $\Sigma$5 &  $\Sigma$5 &  $\Sigma$7 &  $\Sigma$7 &  $\Sigma$9 &  $\Sigma$9\\\hline  
plane &  (011) &  (111) &  (112) &  (001) &  (012) &  (013) &  (111) &  (123) & (011) & (122) \\\hline  
\endfirsthead
\hline
\endhead
\hline
\endfoot
\hline
\endlastfoot
Ag &    1.187 &    1.476 &    1.309 &    1.201 &    1.204 &    1.229 &    1.337 &    1.179 &    1.023 &    1.142 \\\hline Al &    1.498 &    1.595 &    1.649 &    1.447 &    1.498 &    1.512 &    1.456 &    1.427 &    1.242 &    1.466 \\\hline Au &    1.197 &    1.458 &    1.305 &    1.399 &    1.291 &    1.365 &    1.309 &    1.255 &    1.048 &    1.167 \\\hline Ba &    0.578 &    0.592 &    0.721 &    0.364 &    0.535 &    0.534 &    0.563 &    0.582 &    0.526 &    0.559 \\\hline Ca &    0.844 &    0.904 &    0.887 &    0.654 &    0.793 &    0.765 &    0.813 &    0.783 &    0.701 &    0.819 \\\hline Ce &    1.763 &    1.817 &    1.789 &    1.687 &    1.906 &    1.789 &    1.751 &    1.817 &    1.636 &    1.737 \\\hline Cr &    5.727 &    4.890 &    6.402 &    4.619 &    4.884 &    5.069 &    4.536 &    4.994 &    5.225 &    4.424 \\\hline Cs &    0.123 &    0.144 &    0.140 &    0.124 &    0.131 &    0.120 &    0.132 &    0.139 &    0.124 &    0.131 \\\hline Cu &    2.274 &    2.557 &    2.618 &    2.184 &    2.197 &    2.296 &    2.259 &    2.241 &    1.956 &    2.100 \\\hline Fe &    4.386 &    3.862 &    4.794 &    2.756 &    3.241 &    5.308 &    3.556 &    3.846 &    3.924 &    3.557 \\\hline Ir &    3.802 &    4.217 &    3.802 &    4.339 &    3.888 &    4.310 &    3.712 &    3.848 &    3.413 &    3.595 \\\hline K  &    0.185 &    0.176 &    0.228 &    0.155 &    0.181 &    0.162 &    0.195 &    0.180 &    0.193 &    0.183 \\\hline Li &    0.951 &    0.916 &    1.023 &    0.651 &    0.861 &    0.839 &    0.900 &    0.927 &    0.915 &    0.835 \\\hline Mo &    5.092 &    4.181 &    6.318 &    3.931 &    4.170 &    4.448 &    3.715 &    4.252 &    4.555 &    4.066 \\\hline Na &    0.374 &    0.378 &    0.495 &    0.298 &    0.376 &    0.338 &    0.377 &    0.389 &    0.363 &    0.353 \\\hline Nb &    3.854 &    3.392 &    4.439 &    3.059 &    3.248 &    3.366 &    3.335 &    3.540 &    3.550 &    3.347 \\\hline Ni &    3.362 &    3.854 &    3.321 &    3.329 &    3.410 &    3.531 &    3.335 &    3.309 &    2.904 &    3.185 \\\hline Pb &    0.421 &    0.439 &    0.382 &    0.345 &    0.426 &    0.452 &    0.444 &    0.411 &    0.350 &    0.341 \\\hline Pd &    2.262 &    2.605 &    2.545 &    2.333 &    2.249 &    2.370 &    2.357 &    2.222 &    1.959 &    2.141 \\\hline Pt &    2.878 &    2.782 &    2.851 &    3.067 &    2.680 &    2.866 &    2.678 &    2.652 &    2.577 &    2.404 \\\hline Rb &    0.164 &    0.161 &    0.187 &    0.129 &    0.149 &    0.135 &    0.172 &    0.164 &    0.157 &    0.151 \\\hline Rh &    3.109 &    3.896 &    3.370 &    3.480 &    3.318 &    3.541 &    3.324 &    3.272 &    2.807 &    3.138 \\\hline Sr &    0.628 &    0.695 &    0.658 &    0.494 &    0.590 &    0.569 &    0.597 &    0.574 &    0.505 &    0.543 \\\hline Ta &    4.350 &    3.875 &    5.050 &    3.016 &    3.643 &    3.649 &    3.842 &    3.986 &    3.952 &    3.751 \\\hline Th &    1.953 &    2.495 &   -0.000 &    1.895 &    1.879 &    1.814 &    2.285 &    1.820 &    1.729 &    1.866 \\\hline V  &    4.521 &    4.225 &    5.131 &    3.379 &    3.816 &    3.678 &    4.146 &    4.352 &    4.232 &    4.089 \\\hline W  &    5.743 &    4.689 &    6.123 &    4.703 &    4.694 &    5.219 &    4.065 &    4.595 &    5.057 &    4.343 \\\hline Yb &    0.778 &    0.899 &    0.803 &    0.608 &    0.717 &    0.701 &    0.772 &    0.711 &    0.624 &    0.675 \\\hline 
\end{longtable}
\end{center}

\begin{center}
\begin{longtable}[c]{cccccccccc}
\caption{Work of separation ($W_{sep}$) of hcp elements}
\label{SI:Wsep_hcp}\\
\hline
{} &        Be &        Cd &        Co &       Dy &        Er &       Gd&     Hf &        Ho &        La  \\\hline 
$\Sigma$7(0001)  &  2.460 & 0.284 & 3.493 & 1.714 & 1.790 & 1.481 & 3.007 & 1.751 & 1.252  \\\hline 
{} & Lu   & Mg &        Nd &        Os &        Pm &        Pr &        Re &        Ru &        Sc  \\\hline 
$\Sigma$7(0001)& 1.911  &  0.890 &  1.430 &  4.262 &  1.519 &  1.373 &  4.192 &  4.035 &  2.240 \\\hline 

{} &        Sm &      Tb &     Tc &        Ti &        Tl &        Tm &         Y &        Zn &        Zr \\\hline 
$\Sigma$7(0001) & 1.557 &  1.672 &  3.797 &  3.566 &  0.437 &  1.833 &  1.793 &  0.554 &  2.907 {}\\
\hline 
\end{longtable}
\end{center}
\begin{figure}[H]
\subfigure[bcc Mo]
{\label{SI:bcc_Mo_Esurf_GBE_Wsep}\includegraphics[width=0.52\textwidth]{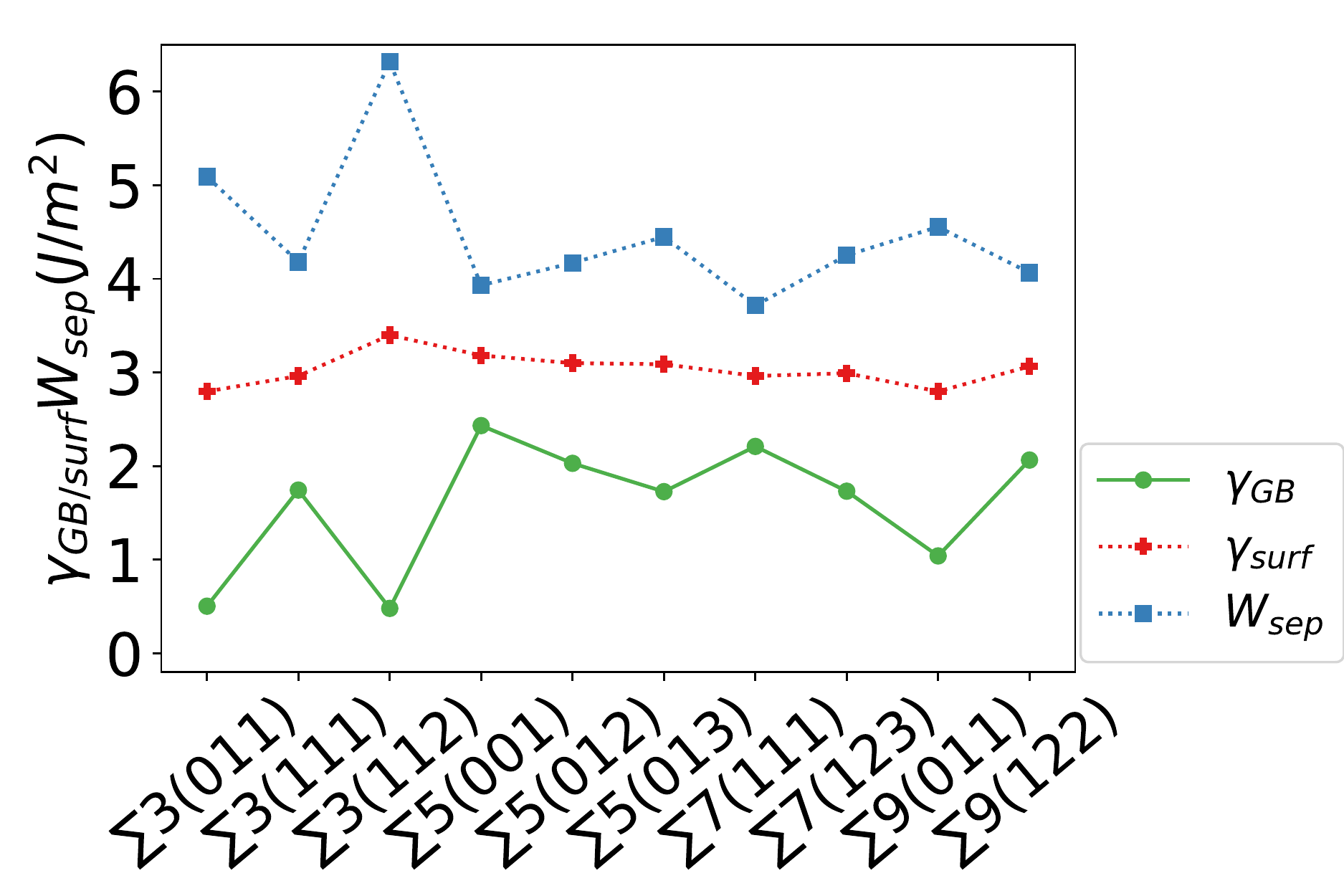}}
\subfigure[fcc Ni]
{\label{SI:fcc_Ni_Esurf_GBE_Wsep}\includegraphics[width=0.52\textwidth]{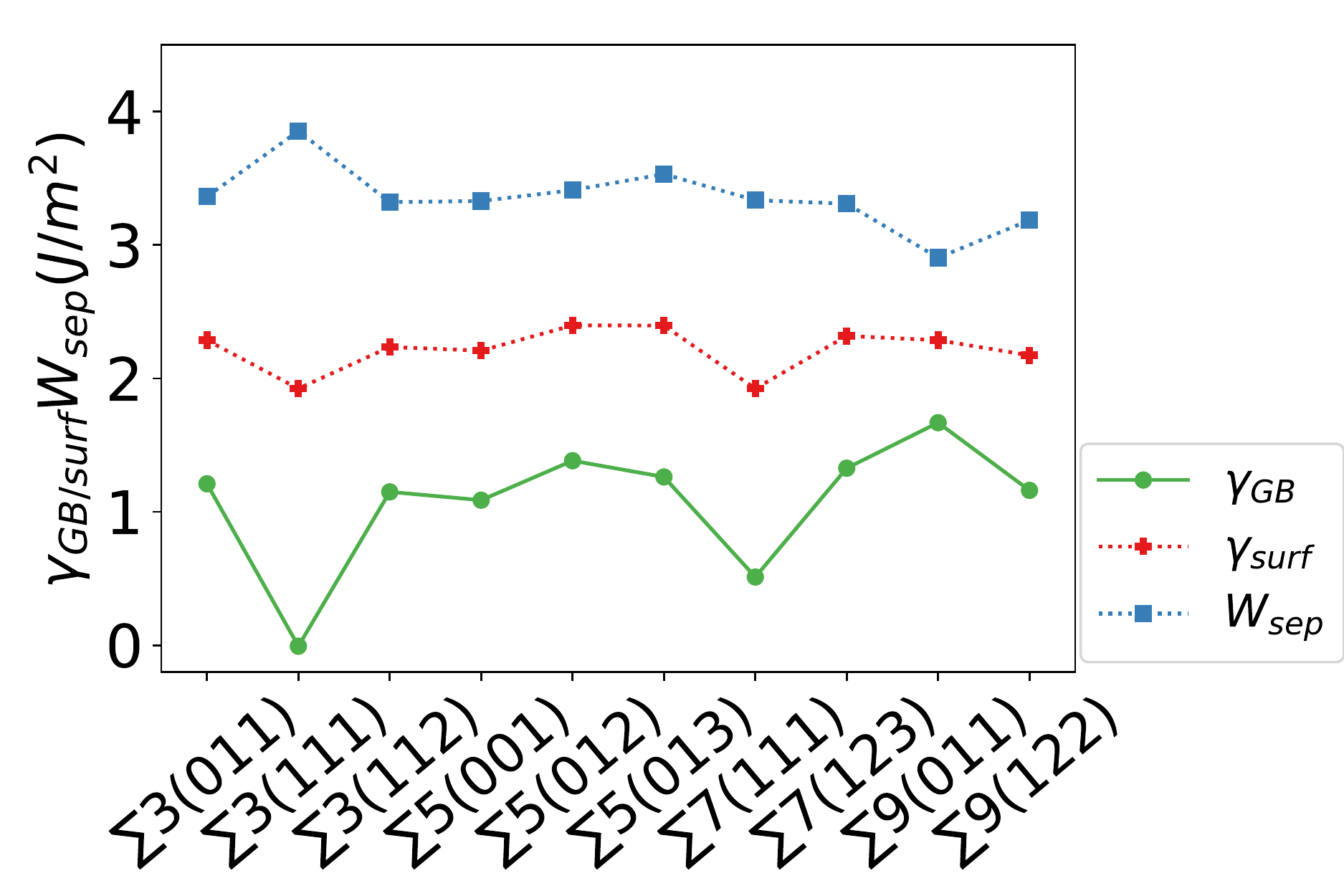}}
\caption{\label{SI:Mo_Ni_E_GB_E_surf_Wsep_vs_GB_types} The $\gamma_{surf}$ varies moderately across different surfaces while the $\gamma_{GB}$ varies dramatically across different GB types. This causes the near-linear relationship between $\gamma_{GB}$ and $W_{sep}$ as shown in Figure S\ref{SI:Wsep_vs_GB_energy}.}
\end{figure}

\begin{figure}[H]
\subfigure[bcc elements]
{\label{SI:bcc_all_GB_energy_vs_Wsep}\includegraphics[width=0.5\textwidth]{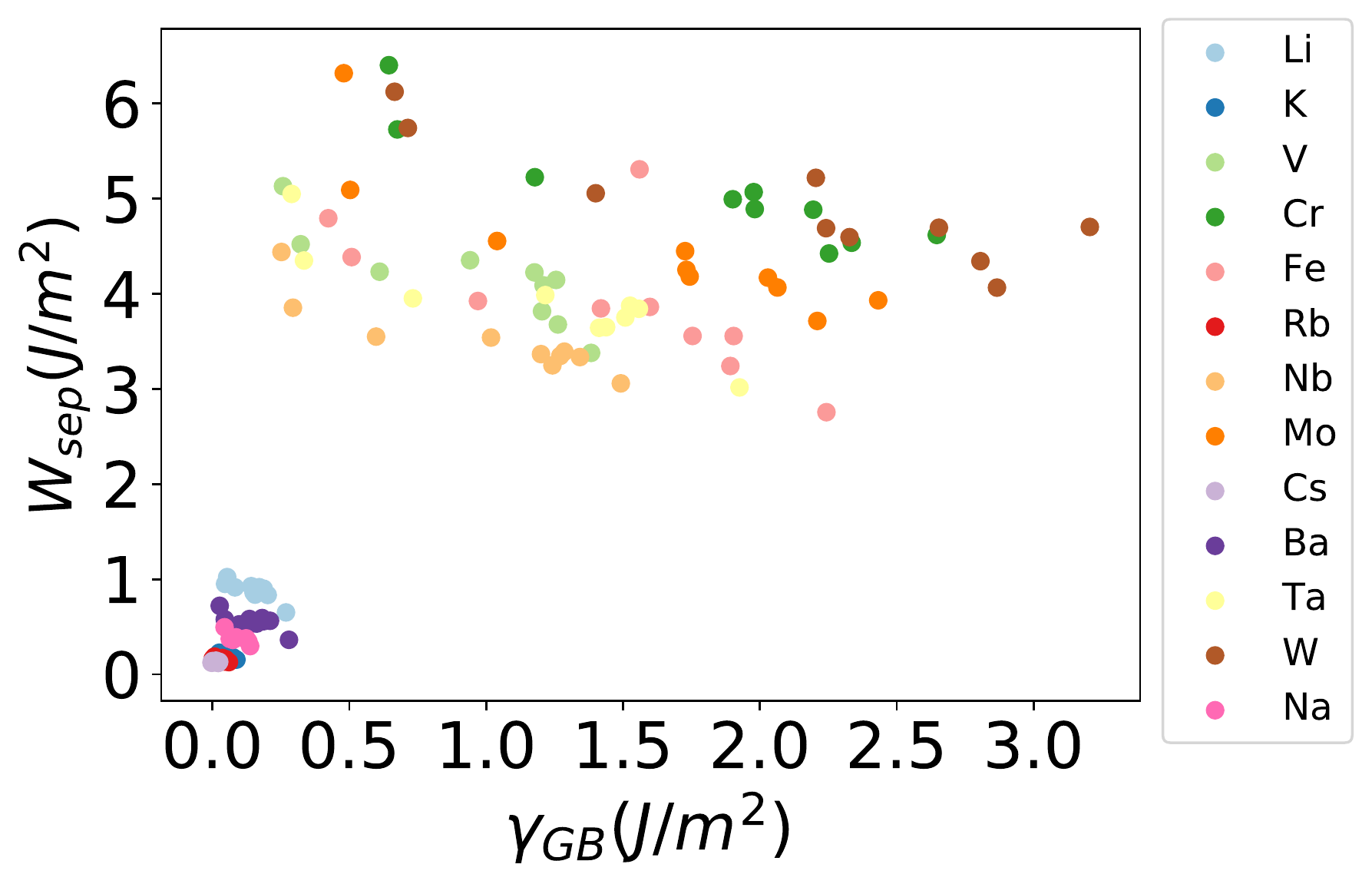}}
\subfigure[fcc elements]
{\label{SI:fcc_all_GB_energy_vs_Wsep}\includegraphics[width=0.5\textwidth]{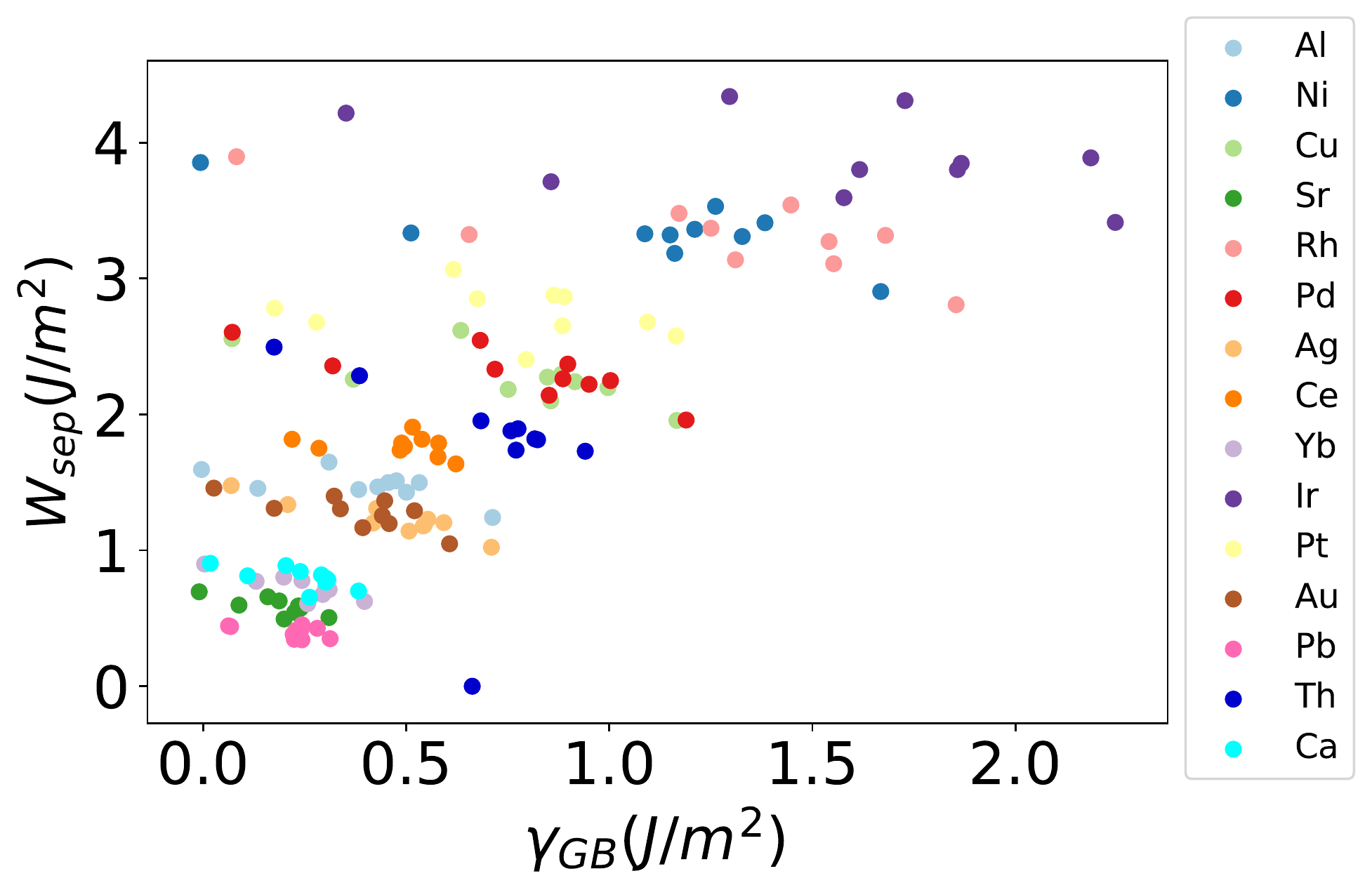}}
\caption{\label{SI:Wsep_vs_GB_energy} The near linear trend between work of separation and GB energy}
\end{figure}

% \newpage
\begin{center}
\begin{longtable}[c]{ccccccc}
\caption{DFT calculated Grain boundary energies from references}
\label{SI:GBenergies_compare_to_other_DFT}\\
\hline
{} & element &          GB type &  This work $\gamma_{GB}$ &  Ref $\gamma_{GB}$ & functional &                      ref. \\\hline 
\endfirsthead
\hline
{} & element &          GB type &  This work $\gamma_{GB}$ &  Ref $\gamma_{GB}$ & functional &                      ref. \\\hline 
\endhead
\hline
\endfoot
\hline
\endlastfoot

0  &      Mo &  $\Sigma$3[1 1 0] (112) &                    0.480 &              0.544 &    PBEsol &     \cite{Scheiber2016f} \\\hline 
1  &      Mo &  $\Sigma$3[1 1 0] (111) &                    1.743 &              1.931 &    PBEsol &     \cite{Scheiber2016f}\\\hline 
2  &      Mo &  $\Sigma$5[1 0 0] (013) &                    1.727 &              1.813 &    PBEsol &     \cite{Scheiber2016f}\\\hline 
3  &      Mo &  $\Sigma$7[1 1 1] (123) &                    1.732 &              1.889 &    PBEsol &     \cite{Scheiber2016f}\\\hline 
4  &      Fe &  $\Sigma$3[1 1 0] (112) &                    0.423 &              0.509 &    PBEsol &     \cite{Scheiber2016f}\\\hline 
5  &      Fe &  $\Sigma$3[1 1 0] (111) &                    1.598 &              1.785 &    PBEsol &     \cite{Scheiber2016f}\\\hline 
6  &      Fe &  $\Sigma$5[1 0 0] (013) &                    1.560 &              1.666 &    PBEsol &     \cite{Scheiber2016f}\\\hline 
7  &      Fe &  $\Sigma$7[1 1 1] (123) &                    1.419 &              1.631 &    PBEsol &     \cite{Scheiber2016f}\\\hline 
8  &       W &  $\Sigma$3[1 1 0] (112) &                    0.665 &              0.655 &    PBEsol &     \cite{Scheiber2016f}\\\hline 
9  &       W &  $\Sigma$3[1 1 0] (111) &                    2.242 &              2.440 &    PBEsol &     \cite{Scheiber2016f}\\\hline 
10 &       W &  $\Sigma$5[1 0 0] (013) &                    2.204 &              2.266 &    PBEsol &    \cite{Scheiber2016f} \\\hline 
11 &       W &  $\Sigma$7[1 1 1] (123) &                    2.327 &              2.371 &    PBEsol &     \cite{Scheiber2016f}\\\hline 
12 &      Fe &  $\Sigma$3[1 1 0] (110) &                    0.508 &              0.520 &    GGA-PBE &         \cite{Wang2018i} \\\hline 
13 &      Fe &  $\Sigma$9[1 1 0] (221) &                    1.754 &              1.660 &    GGA-PBE &         \cite{Wang2018i} \\\hline 
14 &      Fe &  $\Sigma$5[1 0 0] (110) &                    2.243 &              2.120 &    GGA-PBE &         \cite{Wang2018i} \\\hline 
15 &      Fe &  $\Sigma$7[1 1 1] (123) &                    1.419 &              1.460 &    GGA-PBE &         \cite{Wang2018i} \\\hline 
16 &      Fe &  $\Sigma$3[1 1 0] (112) &                    0.423 &              0.450 &    GGA-PBE &         \cite{Wang2018i} \\\hline 
17 &      Fe &  $\Sigma$3[1 1 0] (111) &                    1.598 &              1.570 &    GGA-PBE &         \cite{Wang2018i} \\\hline 
18 &      Fe &  $\Sigma$3[1 0 0] (013) &                    1.560 &              1.570 &    GGA-PBE &         \cite{Wang2018i} \\\hline 
19 &      Fe &  $\Sigma$5[1 0 0] (012) &                    1.892 &              1.640 &    GGA-PBE &         \cite{Wang2018i} \\\hline 
20 &      Fe &  $\Sigma$3[1 1 0] (111) &                    1.598 &              1.610 &    GGA-PBE &  \cite{Bhattacharya2013}\\\hline 
21 &      Fe &  $\Sigma$3[1 1 0] (111) &                    1.598 &              1.570 &   GGA-PW91 &    \cite{Wachowicz2010}\\\hline 
22 &      Fe &  $\Sigma$3[1 1 0] (111) &                    1.598 &              1.520 &        GGA &           \cite{Gao2009} \\\hline 
23 &      Fe &  $\Sigma$3[1 1 0] (112) &                    0.423 &              0.340 &        GGA &           \cite{Gao2009} \\\hline 
24 &      Fe &  $\Sigma$3[1 1 0] (112) &                    0.423 &              0.470 &   GGA-PW91 &           \cite{Du2011} \\\hline 
25 &      Fe &  $\Sigma$5[1 0 0] (013) &                    1.560 &              1.530 &   GGA-PW91 &           \cite{Du2011} \\\hline 
26 &      Fe &  $\Sigma$5[1 0 0] (013) &                    1.560 &              1.630 &    GGA-PBE &           \cite{Cak2008} \\\hline 
27 &      Fe &  $\Sigma$5[1 0 0] (013) &                    1.560 &              1.489 &        GGA &           \cite{Gao2009} \\\hline 
28 &      Fe &  $\Sigma$5[1 0 0] (012) &                    1.892 &              2.000 &   GGA-PW91 &    \cite{Wachowicz2010}\\\hline 
29 &      Ta &  $\Sigma$5[1 0 0] (013) &                    1.438 &              1.544 &       LDFT &         \cite{Ochs2000a} \\\hline 
30 &       W &  $\Sigma$5[1 0 0] (013) &                    2.204 &              2.235 &       LDFT &         \cite{Ochs2000a} \\\hline 
31 &      Mo &  $\Sigma$5[1 0 0] (013) &                    1.727 &              1.700 &       LDFT &         \cite{Ochs2000a} \\\hline 
32 &      Nb &  $\Sigma$5[1 0 0] (013) &                    1.200 &              1.288 &       LDFT &         \cite{Ochs2000a} \\\hline 
33 &      Cu &  $\Sigma$5[1 0 0] (012) &                    0.997 &              0.920 &    GGA-PBE &         \cite{Bean2016a} \\\hline 
34 &      Cu &  $\Sigma$3[1 1 0] (111) &                    0.071 &              0.020 &    GGA-PBE &         \cite{Bean2016a} \\\hline 
35 &      Cu &  $\Sigma$3[1 1 0] (112) &                    0.634 &              0.570 &    GGA-PBE &         \cite{Bean2016a} \\\hline 
36 &      Ni &  $\Sigma$5[1 0 0] (012) &                    1.383 &              1.230 &    GGA-PBE &         \cite{Bean2016a} \\\hline 
37 &      Ni &  $\Sigma$3[1 1 0] (111) &                   -0.007 &              0.040 &    GGA-PBE &         \cite{Bean2016a} \\\hline 
38 &      Ni &  $\Sigma$3[1 1 0] (111) &                    1.150 &              0.840 &    GGA-PBE &         \cite{Bean2016a} \\\hline 
39 &      Al &  $\Sigma$3[1 1 0] (112) &                    0.310 &              0.426 &        LDA &       \cite{Wright1994} \\\hline 

\hline 
\end{longtable}
\end{center}

\newpage
\begin{figure}[H]
\centering\includegraphics[width=0.85\linewidth]{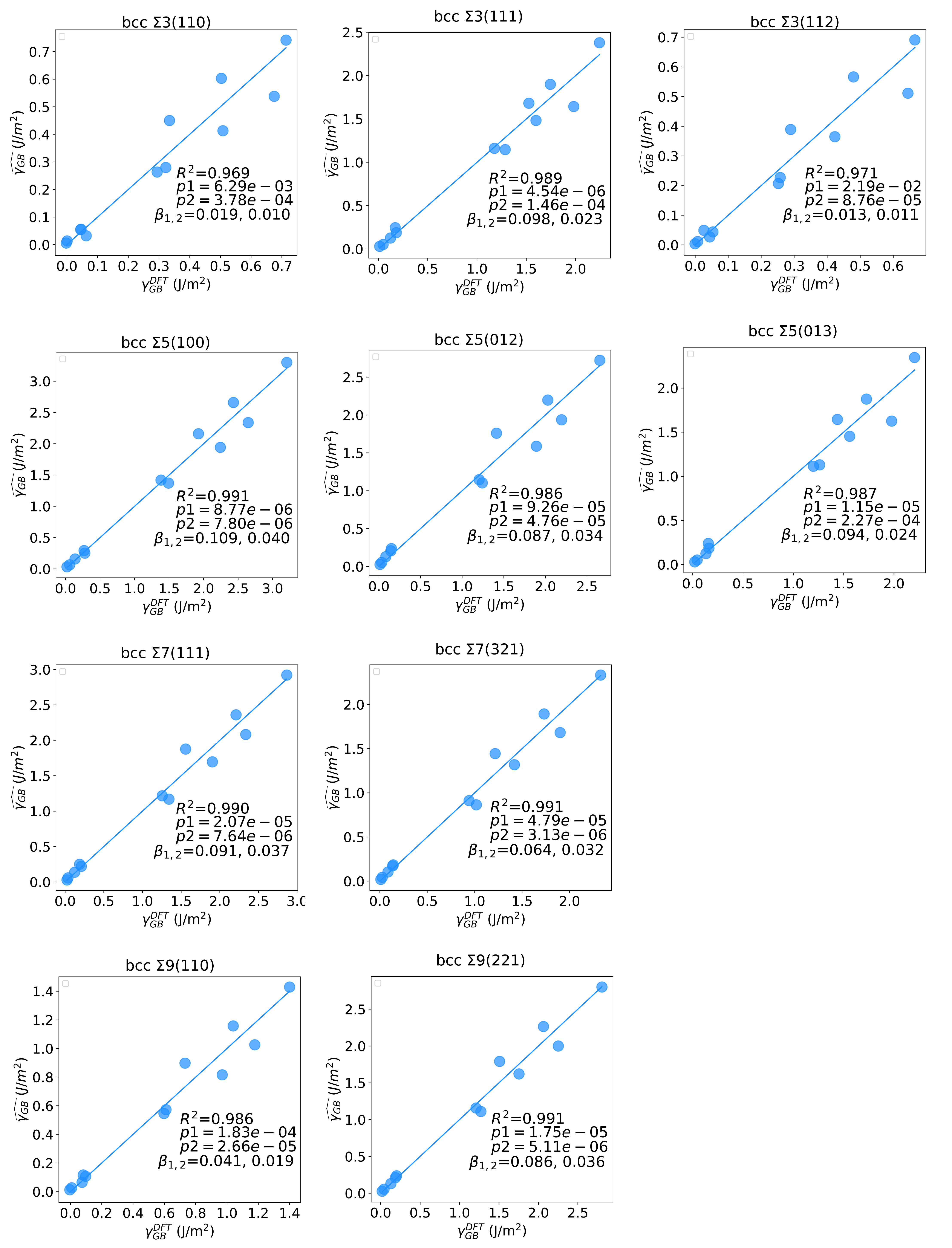}
\caption{\label{SI:bcc_fitting_set} Fitting models for all the GB types of bcc elements with $\gamma_{GB} = \beta_{1}E_{coh}a_0^{-2} + \beta_{2}G\cdot a_0$.}
\end{figure}

\newpage
\begin{figure}[H]
\centering\includegraphics[width=0.85\linewidth]{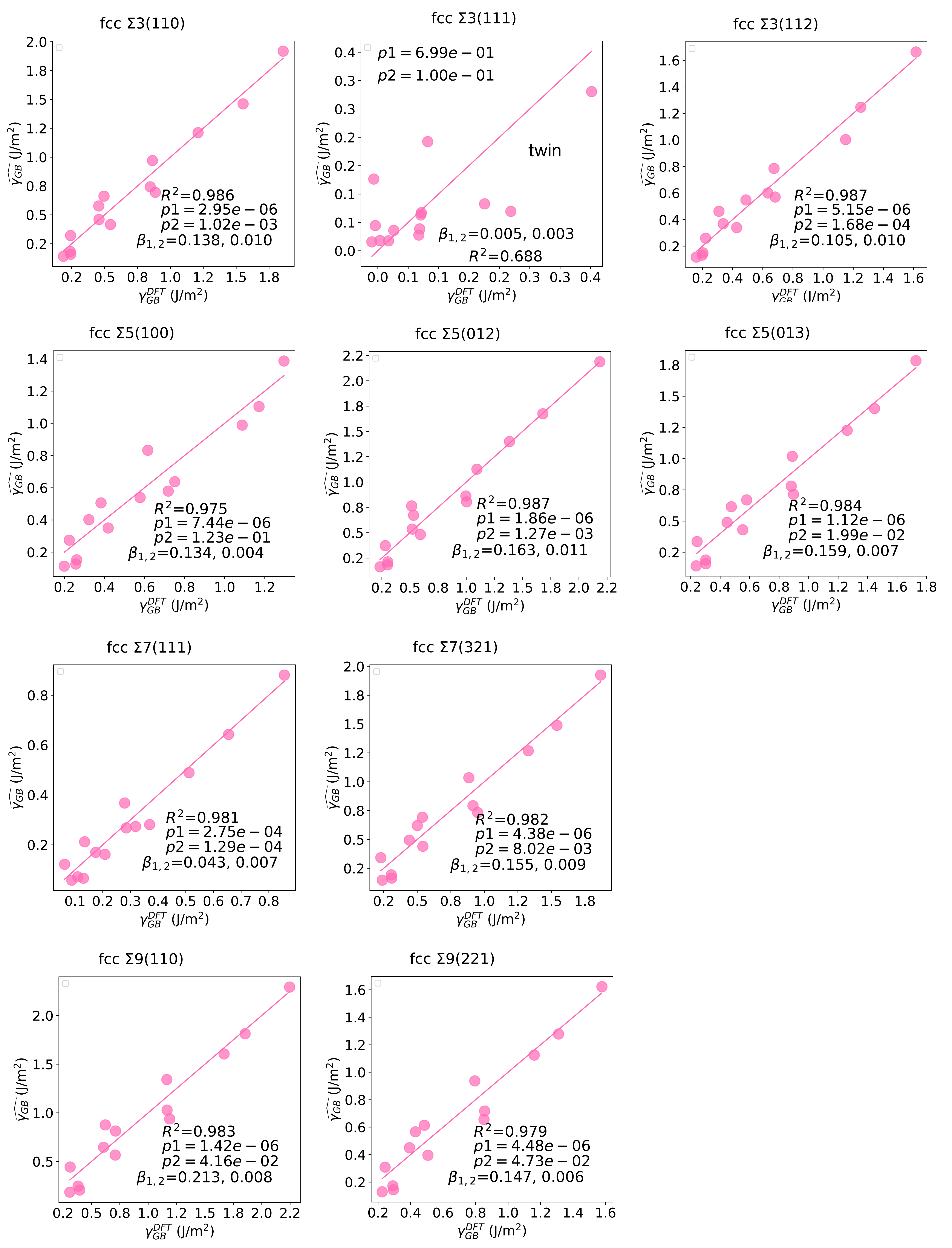}
\caption{\label{SI:fcc_fitting_set} Fitting models for all the GB types of fcc elements with $\gamma_{GB} = \beta_{1}E_{coh}a_0^{-2} + \beta_{2}G\cdot a_0$.}
\end{figure}

% \newpage
\section*{References}
\bibliography{GBDB.bib}